\documentclass[twocolumn,nofootinbib,superscriptaddress]{revtex4-1}

\usepackage{epsfig}
\usepackage{calligra,amsmath,amsfonts,mathrsfs,amssymb,bbm,textcomp,bbding}
\usepackage{slashed,cancel,units}
\usepackage[usenames,dvipsnames]{xcolor}
\usepackage{hyperref}
\hypersetup{colorlinks=true,urlcolor=Magenta,anchorcolor=blue,citecolor=blue,filecolor=blue,
            linkcolor=Magenta,menucolor=blue, linktocpage=true,pdfproducer=medialab}
\usepackage[utf8]{inputenc}

\providecommand{\openone}{\leavevmode\hbox{\small1\kern-3.8pt\normalsize1}}

\newcommand{\sL}{\mathscr{L}}
\newcommand{\sB}{ B}

\newcommand{\sT}{ T}

\newcommand{\nb}{\mathtt{b}}
\newcommand{\nB}{\mathtt{B}}
\newcommand{\nh}{\mathtt{h}}
\newcommand{\nK}{\mathtt{K}}
\newcommand{\nq}{\mathtt{q}}
\newcommand{\nQ}{\mathtt{Q}}
\newcommand{\ns}{\mathtt{s}}
\newcommand{\nt}{\mathtt{t}}
\newcommand{\nT}{\mathtt{T}}

\newcommand{\cB}{\mathcal{B}}
\newcommand{\cG}{\mathcal{G}}
\newcommand{\cH}{\mathcal{H}}
\newcommand{\cM}{\mathcal{M}}

\newcommand{\cQ}{\mathcal{Q}}
\newcommand{\cT}{\mathcal{T}}
\newcommand{\cY}{\mathcal{Y}}
\newcommand{\cV}{\mathcal{V}}

\newcommand{\s}{\sigma}
\newcommand{\derp}{\partial}
\newcommand{\hc}{\text{h.c.}}
\newcommand{\ov}[1]{\overline{#1}}
\newcommand{\nn}{\nonumber}
\newcommand{\vev}[1]{\langle#1\rangle}
\newcommand{\diag}{\mathtt{diag}}
\newcommand{\unity}{\mathbbm{1}}

\newcommand{\TeV}{\ \text{TeV}}
\newcommand{\GeV}{\ \text{GeV}}

\newcommand{\be}{\begin{equation}}
\newcommand{\ee}{\end{equation}}
\def\MLsM{ML$\sigma$M}
\definecolor{vierde}{rgb}{0.0, 0.5, 0.0}

\newcommand{\lag}{\mathcal{L}}

\newcommand{\mo}{M_1}
\newcommand{\mosq}{M^2_1}

\newcommand{\mf}{M_5}
\newcommand{\mfsq}{M^2_5}

\newcommand{\lamo}{\Lambda_1}

\newcommand{\lamt}{\Lambda_2}

\newcommand{\lamth}{\Lambda_3}

\newcommand{\mpf}{M'_5}
\newcommand{\mpfsq}{M'^2_5}

\newcommand{\lampo}{\Lambda'_1}

\newcommand{\lampt}{\Lambda'_2}

\newcommand{\lampth}{\Lambda'_3}

\newcommand{\yo}{\text{y}_1}
\newcommand{\yt}{\text{y}_2}

\newcommand{\mpo}{M'_1}

\newcommand{\ypo}{\text{y}'_1}
\newcommand{\ypt}{\text{y}'_2}

\newcommand{\et}{\mbox{${E_T}$}}
\newcommand{\etmiss}{\mbox{$\protect \raisebox{.3ex}{$\not$}\et$}}

\DeclareMathAlphabet{\mathpzc}{OT1}{pzc}{m}{it}
\begin{document}

\eprint{IFT-UAM/CSIC-19-148\hspace{2cm}FTUAM-19-20}

\title{Exotic Vector-Like Quark Phenomenology in the Minimal Linear $\sigma$ Model}

\author{J. A. Aguilar--Saavedra}
\email{jaas@ugr.es}
\affiliation{Instituto de F\'isica Te\'orica UAM/CSIC, Calle Nicol\'as Cabrera 13-15, Cantoblanco E-28049 Madrid, Spain}
\affiliation{Universidad de Granada, E-18071 Granada, Spain (on leave)}

\author{J.~Alonso-Gonz\'alez}
\email{j.alonso.gonzalez@csic.es}
\affiliation{Instituto de F\'isica Te\'orica UAM/CSIC, Calle Nicol\'as Cabrera 13-15, Cantoblanco E-28049 Madrid, Spain}
\affiliation{Departamento  de  F\'{\i}sica Te\'{o}rica,  Universidad  Aut\'{o}noma  de  Madrid, Cantoblanco  E-28049  Madrid,  Spain}

\author{L.~Merlo}
\email{luca.merlo@uam.es}
\affiliation{Instituto de F\'isica Te\'orica UAM/CSIC, Calle Nicol\'as Cabrera 13-15, Cantoblanco E-28049 Madrid, Spain}
\affiliation{Departamento  de  F\'{\i}sica Te\'{o}rica,  Universidad  Aut\'{o}noma  de  Madrid, Cantoblanco  E-28049  Madrid,  Spain}

\author{J.~M.~No}
\email{josemiguel.no@uam.es}
\affiliation{Instituto de F\'isica Te\'orica UAM/CSIC, Calle Nicol\'as Cabrera 13-15, Cantoblanco E-28049 Madrid, Spain}
\affiliation{Departamento  de  F\'{\i}sica Te\'{o}rica,  Universidad  Aut\'{o}noma  de  Madrid, Cantoblanco  E-28049  Madrid,  Spain}

\begin{abstract}
Extensions of the Standard Model that include vector-like quarks commonly also include additional particles that may mediate new production or decay modes. 
Using as example the minimal linear $\sigma$ model, that reduces to the minimal $SO(5)/SO(4)$ composite Higgs model in a specific limit, we consider the phenomenology of vector-like quarks when a scalar singlet $\sigma$ is present. 
This new particle may be produced in the decays $T \to t \sigma$, $B \to b \sigma$, where $T$ and $B$ are vector-like quarks of charges $2/3$ and $-1/3$, 
respectively, with subsequent decay $\sigma \to W^+ W^-, ZZ, hh$. By scanning over the allowed parameter space  we find that these decays may be dominant. 
In addition, we find that the presence of several new particles allows for single $T$ production cross sections larger than those expected in minimal models. 
We discuss the observability of these new signatures in existing searches.
\end{abstract}

\maketitle

\section{Introduction}
\label{sec:intro}

The electroweak (EW) hierarchy problem remains one of the weaknesses of Standard Model (SM) of particle physics. Indeed, a potentially strong fine tuning 
in the scalar sector is required once the need for heavy new physics is considered, for example to explain non-vanishing neutrino masses and the presence of dark 
matter. Among the best known scenarios to solve this problem, the composite Higgs (CH) framework~\cite{Kaplan:1983fs,Kaplan:1983sm,Banks:1984gj} has received 
a renewed interest~\cite{Agashe:2004rs,Barbieri:2007bh,Gripaios:2009pe,Mrazek:2011iu}. The traditional CH model is based on a global symmetry $\cG$, broken 
spontaneously to a subgroup $\cH$, such that the coset $\cG/\cH$ is symmetric. A certain number of Nambu-Goldstone bosons (GBs) arise from this breaking: 
in particular, the SM GBs, which are the would-be longitudinal components of the EW gauge bosons, and the Higgs itself. This technically solves the hierarchy 
problem, providing a symmetry protection for the Higgs mass.

The idea is that an undetermined strong dynamics, acting at a high-scale, generates exotic fermionic bound states that spontaneously break the 
group $\cG$. Moreover, the gauging of the SM symmetries, together with non-universal fermion masses, provide an explicit breaking 
of  $\cG$ such that a small mass is generated for the Higgs. SM fermion masses are generated by means of the fermion partial compositeness 
mechanism~\cite{Dugan:1984hq,Kaplan:1991dc,Contino:2004vy}, which consists in introducing in the spectrum a series of exotic (typically
vector-like) fermions which act as partners of the SM fermions, with a role similar to that of right-handed (RH) Majorana neutrinos in the Type I Seesaw 
mechanism~\cite{Minkowski:1977sc,Weinberg:1979sa,Yanagida:1980xy} for neutrino masses. In the mass basis, this results in light SM fermions and heavier 
exotic counterparts. The top quark is the heaviest among the light fermions and it turns out to be largely composed of exotic states. 
On the other hand, the lightest exotic fermions are the counterparts of the top and bottom quarks. These are labelled, depending on their electric 
charge, as top-partner $\sT$ and bottom-partner $\sB$, respectively. 
The search for heavy partners of the SM quarks is a very effervescent field from the experimental 
side~\cite{Aaboud:2017qpr,Aaboud:2017zfn,Sirunyan:2017pks,Sirunyan:2018fjh,Aaboud:2018xuw,Sirunyan:2018omb,Aaboud:2018uek,Aaboud:2018saj,
Aaboud:2018xpj,Aaboud:2018wxv,Sirunyan:2019tib,Sirunyan:2018yun,Aaboud:2018ifs,Aaboud:2018zpr,Sirunyan:2018qau,Sirunyan:2019sza,Sirunyan:2019xeh}. 
The discovery of such particles would not only represent a fundamental step toward the understanding of the theory beyond the SM, but also a 
window to study the electroweak symmetry breaking (EWSB) sector.

In this paper the focus is in the so-called Minimal Linear $\sigma$ Model (\MLsM)~\cite{Feruglio:2016zvt}, that attempts to improve the 
limitations of the traditional CH models. In the latter, the indetermination of the strong dynamics leads to describe the model 
Lagrangian with an effective approach, which however leads to a limited range of application. The \MLsM, instead, is a renormalisable 
model based on the global spontaneous symmetry breaking $SO(5)\to SO(4)$ that matches the minimal 
CH model~\cite{Agashe:2004rs,Alonso:2014wta,Panico:2015jxa,Hierro:2015nna} in a specific limit. The main 
idea is the introduction of a scalar quintuplet of $SO(5)$ that contains as degrees of freedom the three SM GBs, 
the Higgs and an additional EW singlet $\sigma$. Moreover, the spectrum is enlarged by the introduction of several 
exotic fermions in the trivial and fundamental representations of $SO(5)$ that will give rise to the fermion partial compositeness mechanism. 

Several studies have been conducted on this model to analyse its features at low-energies~\cite{Gavela:2016vte}, 
projecting to the so-called Higgs effective field theory Lagrangian~\cite{Feruglio:1992wf,Grinstein:2007iv,Contino:2010mh,Alonso:2012px,Alonso:2012pz,Buchalla:2013rka,
Brivio:2013pma,Brivio:2014pfa,Gavela:2014vra,Gavela:2016bzc,Eboli:2016kko,Brivio:2016fzo,Merlo:2016prs,Alonso:2017tdy}, to access 
its ability to solve the strong CP problem~\cite{Brivio:2017sdm,Merlo:2017sun,Alonso-Gonzalez:2018vpc}. The focus of this 
paper, instead, is to study the phenomenology associated to the exotic vector-like quarks (VLQs). Indeed, being a renormalisable model, all the parameters describing the low-energy theory are fixed in 
terms of the original Lagrangian parameters. 
In particular, the whole spectrum of the exotic VLQs and their interactions are fixed in terms of the initial parameters: this is a 
novelty with respect to previous studies where no direct link was present between the masses of the different VLQs, neither with their 
interactions, which were taken to be independent parameters.

The new VLQ states can be produced in pairs via QCD interactions in hadron collisions, with cross sections that only depend on 
the VLQ mass. In addition, it is also possible to singly produce the VLQs at colliders via their mixing with SM quarks. In both cases, these cross 
sections quickly decrease as the VLQ mass increases, and therefore for collider phenomenology the most interesting signals 
are generically those of the lightest VLQ.\footnote{Since single VLQ production also depends on the fermion mixing, the highest production cross section may not always correspond to the lightest VLQ.}
The electric charge of the lightest VLQ (and 
therefore its decay modes) depends on the model parameters. In case it has exotic charge $5/3$ or $-4/3$, its phenomenology is quite the 
same as in minimal models~\cite{AguilarSaavedra:2009es,Aguilar-Saavedra:2013qpa}, because it does not couple to the scalar $\sigma$. On 
the other hand, if it has charge $2/3$ ($\sT$) or charge $-1/3$ ($\sB$), these VLQ states inherit new decay modes beyond the 
standard decays 
\be
\begin{aligned}
& T \to W b \,,\quad T \to Z t \,,\quad T \to h t \,, \\
& B \to W t \,,\quad B \to Z b \,,\quad B \to h b \,, 
\end{aligned}
\label{ec:TBstd}
\ee
due to the presence of the singlet scalar $\sigma$. These new decays
\be
\begin{aligned}
T \to \sigma t \,\,\,,\;\;\; B \to \sigma b
\end{aligned}
\label{ec:TBexo}
\ee
with
\be
\begin{aligned}
\sigma \to W^+ W^- \,,\; ZZ \,,\; hh
\end{aligned}
\label{ec:TBexo2}
\ee
open a plethora of new possible signatures for VLQs, both in the pair and the single production channels. Current searches, while targeting the standard decays in Eq.~\eqref{ec:TBstd}, are sensitive to the new decays to a varying degree.

Non-standard VLQ top and bottom partner decays into a singlet scalar field like those in Eq.~\eqref{ec:TBexo} 
have already been explored in the literature~\cite{Cheng:2005as,Chala:2017xgc,Dobrescu:2016pda,Benbrik:2019zdp,Bizot:2018tds} 
(see also~\cite{DeSimone:2012fs,Aguilar-Saavedra:2017giu}). These studies mainly consider the singlet scalar $\sigma$ being invisible at 
colliders~\cite{Cheng:2005as,Chala:2017xgc,Colucci:2018vxz}, 
or decaying into $Z\gamma$, $\gamma\gamma$ final states~\cite{Benbrik:2019zdp}, yet not focusing on the singlet 
scalar decays from Eq.~\eqref{ec:TBexo2}, which generically are the main decay modes 
for a singlet scalar $\sigma$ mixing with the SM Higgs boson and with mass $m_{\sigma} >$ 200 GeV.
Moreover, the singlet scalar $\sigma$ has previously been considered to be a GB of the coset $\cG/\cH$ (together with the Higgs boson), 
while in the \MLsM~the nature of the $\sigma$ field is qualitatively very different from that of the Higgs 
(e.g.~$\sigma$ is naively expected to be significantly heavier that the Higgs boson).

\vspace{2mm}

The paper is organised as follows. Section~\ref{sec:MLsM} describes the relevant aspects of the model, writing down the interaction 
Lagrangians that determine the phenomenology of the VLQs. The 
constraints on the parameter space of the model arising either from precision electroweak data, Higgs boson measurements, and direct 
searches, are discussed in Sec.~\ref{sec:Constraints}. In Sec.~\ref{sec:scan} we outline the procedure used for the exploration of the \MLsM~parameter 
space. The results of our parameter scan are analyzed in Sec.~\ref{sec:Phem}, focusing on the phenomenology of the lightest VLQ, 
the presence of new decay modes and opportunities for single VLQ production at the LHC. We also include 
a discussion on how these new decay modes can be experimentally searched for. We finally conclude in Sec.~\ref{sec:5}.

\boldmath
\section{The Minimal Linear $\sigma$ Model}
\unboldmath
\label{sec:MLsM}

The \MLsM\ is based on the group $SO(5)\times U(1)_X$, where the last factor ensures the correct hypercharge assignments for the SM fields. The 
spectrum contains the four SM gauge bosons associated to the SM gauge symmetry, a real scalar field $\phi$ in the fundamental representation 
of $SO(5)$, the elementary fermions with the same quantum numbers as in the SM and finally exotic vector-like quarks in the trivial and fundamental 
representations of $SO(5)$.  

The scalar field $\phi$ includes the three would-be-longitudinal components of the SM gauge bosons $\pi_i$, $i=1,\,2,\,3$, the Higgs field $\nh$ and the 
additional scalar field $\ns$, singlet under the SM group:
\be
\phi=(\pi_1,\,\pi_2,\,\pi_3,\,\nh,\,\ns)^T
\xrightarrow{u.g.}
\phi=\left(0,\,0,\,0,\,\nh,\,\ns\right)^T\,,
\label{phiDefinition}
\ee
where the last expression holds in the unitary gauge (but in the unbroken phase). The associated scalar potential is responsible for the 
spontaneous $SO(5)\to SO(4)$ breaking. As discussed in the following, explicit breaking terms are also contained in the scalar potential and 
are responsible for the $SO(4)$ and EW breaking.

The elementary fermions do not directly couple to the scalar $\phi$, and therefore neither to the SM Higgs once $SO(5)$ is broken. 
These couplings only arise through the mediation of the exotic fermions that do have tree level interactions with $\phi$. The proto-Yukawas are 
couplings between the $SO(5)$ quintuplet VLQs $\psi$ and the singlets $\chi$: these fields have $U(1)_X$ charge equal to $2/3$ ($-1/3$) and are 
associated to the up-type (down-type) sector.

Table~\ref{tab:SO5Transformations} reports the fields in the spectrum together with their transformation properties under $SO(5)\times U(1)_X$. 
While in full generality one set of VLQs can be introduced per SM generation, in this paper only the third generation SM fermions and 
their VLQ siblings are considered, accordingly to the description of the original publication~\cite{Feruglio:2016zvt}. This does not represent 
a restriction in the present analysis, because the top- and bottom-partners are indeed the lightest VLQs and 
therefore they are the first exotic states 
that may show up in experiments, as discussed before. Moreover, their contributions are the only numerically relevant ones and for this reason 
the tree-level couplings of the first two generation quarks are taken to be purely SM like (deviations from the SM can however be induced at loop-level and will be discussed in Sec.~\ref{sec:Constraints}).

In the remainder of this section, the scalar and fermionic sectors will be discussed following Ref.~\cite{Feruglio:2016zvt}, 
fixing the notation and the conventions that will be used in the rest of the paper. 

\begin{table}[t]
\centering{
\begin{tabular}{|c||c||c|c|c|c|}
\hline
&&&&&\\[-3mm]
		& $\phi$ 	&  $\psi^{(2/3)}$ 	& $\chi^{(2/3)}$ 	& $\psi^{(-1/3)}$ 	& $\chi^{(-1/3)}$  \\[1mm]
\hline
&&&&&\\[-3mm]
$SO(5)$ 	& $5$	& $5$ 		& $1$ 		& $5$ 		& $1$ \\
$U(1)_X$	& $0$ 	& $+2/3$ 		& $+2/3$ 		& $-1/3$ 		& $-1/3$ \\[1mm]
\hline
\end{tabular}
\caption{\em Transformation properties of the fields in the spectrum under $SO(5)\times U(1)_X$. 
The superscripts $(2/3)$ and $(-1/3)$ on the fermionic fields refer to the top and bottom quark sectors.}
\label{tab:SO5Transformations}
}
\end{table}

\subsection{The Scalar Sector}
\label{SubSect:Scalar}

The part of the Lagrangian describing the interactions of the scalar fields and the symmetry breaking is 
\be
\sL_\phi=\dfrac12\left(D_\mu\phi\right)^T\left(D^\mu\phi\right)-V(\phi)\,,
\ee
where the covariant derivative containing the $SU(2)_L\times U(1)_Y$ gauge bosons is defined by
\be
D_\mu\phi=\left(\derp_\mu+ig\,\Sigma_L^a\,W^a_\mu+ig'\,\Sigma_R^3\,B_\mu\right)\phi\,,
\ee
where $\Sigma_L^a$ and $\Sigma_R^a$ denote the generators of $SU(2)_L\times SU(2)_R$, isomorphic of $SO(4)'$, subgroup of $SO(5)$, and rotated with respect to 
the $SO(4)$ residual group after the spontaneous breaking of $SO(5)$. 
The scalar potential $V(\phi)$ contains terms responsible for the spontaneous breaking of $SO(5)$ down to $SO(4)$, plus additional terms that break explicitly 
the residual $SO(4)$,
\be
V(\phi)=\lambda\left(\phi^T\phi-f^2\right)^2+\alpha\,f^3\,\ns-\beta\,f^2\,\nh^2\,,
\label{ScalarPotential}
\ee
where $f$ is the scale at which the $SO(5)$ breaking takes place. The last two terms are just a subset of all the soft breaking terms, but are the ones 
necessary to absorbe divergencies once the one-loop Coleman-Weinberg contributions are considered (see Ref.~\cite{Merlo:2017sun} for a different treatment). 
For values $\lambda\gg1$, the non-linear model would be recovered, corresponding to a decoupling of the $\s$ field.

The $SO(4)$ and EW breaking then require $\nh$ and $\ns$ to develop a vacuum expectation value (VEV)
\be
\vev{\nh}=v_\nh\,,\quad
\vev{\ns}=v_\ns\,,
\label{VEVS}
\ee
where the normalisation has been chosen to match Eq.~(\ref{phiDefinition}), note in particular that $v_\nh = 246$ GeV.
For $\alpha\neq0\neq\beta$, the VEVs turn out to be
\be
\label{vevs_hs}
v_\ns^2=f^2\dfrac{\alpha^2}{4\beta^2}\,,\quad
v_\nh^2=f^2\left(1-\dfrac{\alpha^2}{4\beta^2}+\dfrac{\beta}{2\lambda}\right)\,,
\ee
undergoing the condition
\be
v_\nh^2+v_\ns^2=f^2\left(1+\dfrac{\beta}{2\lambda}\right)\,.
\ee
Requiring the $SO(5)$ breaking ($f^2>0$) and that the Higgs arises as a GB ($\left| v_\nh \right|<\left| v_\ns \right|$) imply that
\be
2\beta^2\left(1+\frac{\beta}{2\lambda}\right)<\alpha^2<4\beta^2\left(1+\frac{\beta}{2\lambda}\right)
\label{TheoContraintsab}
\ee
which in the strongly interacting limit, $\alpha, \beta\ll\lambda$, reduces to 
\be
2\beta^2\leq\alpha^2\leq4\beta^2\,.
\label{TheoContraintsab2}
\ee
Then, in order to get $v_\nh^2\ll f^2$ from Eq.~\eqref{vevs_hs}, $\alpha/2\beta\sim1$ is needed.

Once the scalar fields develop their VEVs, a non diagonal $2\times2$ mass matrix results from Eq.~\eqref{ScalarPotential}. 
Diagonalising this mass matrix, the mass eigenstates $h$, $\sigma$ turn out to be
\be
h=\nh\,\cos\gamma-\ns\sin\gamma\,,\quad
\s=\nh\,\sin\gamma+\ns\cos\gamma\,,
\label{PhysicalScalars}
\ee
where the mixing angle is given by
\be
\tan 2\gamma=\dfrac{4\,v_\nh\,v_\ns}{3v_\ns^2-v_\nh^2-f^2}\,,
\ee
and the masses of these two states, in the limit of $\alpha,\beta\ll\lambda$ and positive $\beta$, read
\be
m^2_\s\simeq8\,\lambda\,f^2+2\beta\,(3f^2-v_\nh^2)\,,\quad
m^2_h\simeq2\,\beta\,v_\nh^2\,.
\label{TheoContraintsBeta}
\ee
As a concluding remark, the possibility $m_\s<m_h$ is viable but extremely fine-tuned and therefore will not be considered. 
As a consequence, the angle $\gamma$ will always be taken in the interval $[-\pi/4,\,\pi/4]$.\\

In the interaction basis, only the SM GBs and $\nh$ couple to the EW gauge bosons. Due to the rotation 
in Eq.~\eqref{PhysicalScalars}, $\s$ inherits couplings with $W^\pm$ and $Z$ weighted by $\sin\gamma$, while the $h$ couplings acquire a suppression with $\cos\gamma$,
\be
\begin{aligned}
\sL_\phi \supset &
-\lambda\Big(h^2+\s^2+2h\s\Big)^2+ \\
&-4\lambda\left(v_\nh\cos\gamma-v_\ns\sin\gamma\right)\Big(h^3+h\s^2\Big)+\\
&-4\lambda\left(v_\nh\sin\gamma+v_\ns\cos\gamma\right)\Big(\s^3+\s h^2\Big)+\\
&+ \left(1+\dfrac{h}{v_\nh}\cos\gamma+\dfrac{\s}{v_\nh}\sin\gamma\right)\times \\
&\qquad\quad\times\left(M_W^2\,W_\mu^+W^{\mu-}+\dfrac12M_Z^2\,Z_\mu Z^\mu\right)\,.
\end{aligned}
\label{eq:lag_s_ints}
\ee
The modification in the $h$ couplings with respect to SM predictions constrain the possible values of $\gamma$, 
as it will be seen in Section~\ref{sec:Constraints}.

\subsection{The Fermionic Sector}

We now turn to discuss the fermionic sector of the \MLsM.
In the basis of canonical kinetic terms for the gauge fields, the renormalisable fermionic Lagrangian can be written as follows~\cite{Feruglio:2016zvt}
\begin{widetext}
\begin{eqnarray}
\sL_f & = & \ov{\nq}_L\,i\slashed{D}\,\nq_L + \ov{\nt}_R\,i\slashed{D}\,\nt_R + \ov{b}_R\,i\slashed{D}\,\nb_R \notag  \\ 
& & +\ov{\psi}^{(2/3)}\left[i\slashed{D}-\mf\right]\psi^{(2/3)}+\ov{\chi}^{(2/3)}\left[i\slashed{D}-\mo\right]\chi^{(2/3)}
+ \ov{\psi}^{(-1/3)}\left[i\slashed{D}-\mpf\right]\psi^{(-1/3)}+\ov{\chi}^{(-1/3)}\left[i\slashed{D}-\mpo\right]\chi^{(-1/3)} \notag \\
& & -\Big[\yo\ov{\psi}^{(2/3)}_L\phi\chi^{(2/3)}_R+\yt\ov{\psi}^{(2/3)}_R\phi\chi^{(2/3)}_L+
\ypo\ov{\psi}^{(-1/3)}_L\phi\chi^{(-1/3)}_R+\ypt\ov{\psi}^{(-1/3)}_R\phi\chi^{(-1/3)}_L 
\label{eq:lag_f} \\ 
& & +\lamo\left(\ov{\nq}_L\Delta^{(2/3)}_{2\times5}\psi^{(2/3)}_R\right)
+\lamt\ov{\psi}^{(2/3)}_L\left(\Delta^{(2/3)}_{5\times1}\nt_R\right)
+\lamth\ov{\chi}^{(2/3)}_L\nt_R 
+\lampo\left(\ov{\nq}_L\Delta^{(-1/3)}_{2\times5}\psi^{(-1/3)}_R\right)
\notag \\ 
& & 
+\lampt\ov{\psi}^{(-1/3)}_L\left(\Delta^{(-1/3)}_{5\times1}\nb_R\right)
+\lampth\ov{\chi}^{(-1/3)}_L\nb_R
+\hc\Big]\,.\nn 
\end{eqnarray}
\end{widetext}
where the $U(1)_X $ charge of the exotic fermion fields $\psi$ and $\chi$ has been made explicit throughout. The first 
line contains the canonical kinetic terms for the elementary quarks, $\nq_L$ for the left-handed (LH) $SU(2)_L$-doublet, $\nt_R$ and $\nb_R$ for the RH $SU(2)_L$ singlets. 
The second line describes the kinetic and mass terms for the exotic quarks. The proto-Yukawa interactions among the exotic fermions and the scalar quintuplet are written 
in the third line. Finally the last two lines show the $SO(5)$ breaking interactions between the exotic and the elementary quarks: the terms $\Delta_{2\times5}$ and 
$\Delta_{5\times1}$ denote spurion fields~\cite{DAmbrosio:2002vsn,Cirigliano:2005ck,Davidson:2006bd,Alonso:2011jd,Dinh:2017smk} that connect the exotic and 
the elementary sector and are the responsible of the light fermion masses. According to the fermion partial compositeness paradigm, no direct elementary 
fermion couplings to $\phi$ are allowed. All the parameters in the previous Lagrangian are taken to be real: this is equivalent to assuming CP 
conservation in these interactions; as only the third generation of light fermions are considered and therefore the CKM matrix cannot be described, this hypothesis is viable.

In terms of the $SU(2)_L$ quantum numbers, the scalar quintuplet $\phi$ and the VLQs can be decomposed as follows
\be
\begin{gathered}
\phi=(H^T,\,\widetilde{H}^T,\,\ns)\,, \\
\psi^{(2/3)}\sim\left(\nK,\,\nQ,\,\nT_5\right)^T\,,\qquad  \chi^{(2/3)}\sim \nT_1\,, \\
\psi^{(-1/3)}\sim\left(\nQ',\,\nK',\,\nB_5\right)^T, \qquad \chi^{(-1/3)}\sim \nB_1\,,
\label{PsiChiComponents}
\end{gathered}
\ee
where $H$ is the SM $SU(2)_L$ doublet, with $\widetilde{H}\equiv i\sigma_2H^\ast$, and $\nK^{(\prime)}$ and $\nQ^{(\prime)}$ are $SU(2)_L$ 
doublets and $\nT_{1,5}$ and $\nB_{1,5}$ are singlets. The rest of the charge assignments can be seen in Table~\ref{tab:SMTransformations}, where the 
hypercharge follows from the relation
\begin{equation}
Y=\Sigma_R^{(3)}+X \,,
\end{equation}
with $X$ the $U(1)_X$ charge and $\Sigma_R^{(3)}$ the third component of the global $SU(2)_R$, which is part of the residual $SO(4)$ group after the breaking of $SO(5)$.

\begin{table*}[htb]
\centering 
\renewcommand{\arraystretch}{1.5}
\footnotesize
\begin{tabular}{|c|c|c|c|c|c|c| }
\hline
Charge/Field & $\nK$ & $\nQ$ & $\nT_{1,5}$ & $\nQ'$ & $\nK'$ & $\nB_{1,5}$ \\[0.5ex] 
\hline
$\Sigma^{(3)}_R$ & $+1/2$ & $-1/2$ & 0 & $+1/2$ & $-1/2$ & 0 \\
\hline
$SU(2)_L \times U(1)_Y$ & $(2,+7/6)$ & $(2,+1/6)$ & $(1,+2/3)$ & $(2,+1/6)$ & $(2,-5/6)$ & $(1,-1/3)$ \\
\hline 
$U(1)_X$ & $+2/3$ & $+2/3$ & $+2/3$ & $-1/3$ & $-1/3$ & $-1/3$ \\
\hline
$U(1)_{EM}$ & 
$\begin{matrix} \nK^u = +5/3 \\ \nK^d = +2/3 \end{matrix}$ & 
$\begin{matrix} \nQ^u = +2/3 \\ \nQ^d = -1/3 \end{matrix}$ & 
$+2/3$ & 
$\begin{matrix} \nQ'^u = +2/3 \\ \nQ'^d = -1/3 \end{matrix}$ & 
$\begin{matrix} \nK'^u = -1/3 \\  \nK'^d = -4/3 \end{matrix}$ & 
$-1/3$ \\
\hline 
\end{tabular}
\caption{\em Decompositions of the exotic fields and their transformations under the SM group.}
\label{tab:SMTransformations}
\end{table*}

In terms of the $SU(2)_L$ components, the fermionic Lagrangian acquires the following form
\begin{widetext}
\be
\begin{aligned}
\hspace{-7mm}
\sL_f=&\phantom{+}\ov{\nq}_L\,i\slashed{D}\,\nq_L+\ov{\nt}_R\,i\slashed{D}\,\nt_R+\ov{\nb}_R\,i\slashed{D}\,\nb_R+\\
&+\ov{\nK}\left[i\slashed{D}-\mf\right]\nK+\ov{\nQ}\left[i\slashed{D}-\mf\right]\nQ
+\ov{\nT}_5\left[i\slashed{D}-\mf\right]\nT_5+\ov{\nT}_1\left[i\slashed{D}-\mo\right]\nT_1+\\ 
&+\ov{\nQ'}\left[i\slashed{D}-\mpf\right]\nQ'+\ov{\nK'}\left[i\slashed{D}-\mpf\right]\nK'
+\ov{\nB}_5\left[i\slashed{D}-\mpf\right]\nB_5+\ov{\nB}_1\left[i\slashed{D}-\mpo\right]\nB_1+\\ 
&-\Big[\yo\left(\ov{\nK}_L\,H\,\nT_{1,R}+\ov{\nQ}_L\,\tilde{H}\,\nT_{1,R}+\ov{\nT}_{5,L}\,\ns\, \nT_{1,R}\right)
+\yt\left(\ov{\nT}_{1,L}\,H^\dagger\, \nK_R+\ov{\nT}_{1,L}\,\tilde{H}^\dagger\, \nQ_R+\ov{\nT}_{1,L}\,\ns\, \nT_{5,R}\right)+\\ 
&\hspace{5mm}+\ypo\left(\ov{\nQ'}_L\,H\,\nB_{1,R}+\ov{\nK'}_L\,\tilde{H}\,\nB_{1,R}+\ov{\nB}_{5,L}\,\ns\, \nB_{1,R}\right)
+\ypt\left(\ov{\nB}_{1,L}\,H^\dagger\, \nQ'_R+\ov{\nB}_{1,L}\,\tilde{H}^\dagger\, \nK'_R+\ov{\nB}_{1,L}\,\ns\, \nB_{5,R}\right)+\\ 
&\hspace{5mm}+\lamo\ov{\nq}_L\nQ_R
+\lamt\ov{\nT}_{5,L}\nt_R
+\lamth\ov{\nT}_{1,L}\nt_R
+\lampo\ov{\nq}_L\nQ'_R
+\lampt\ov{\nB}_{5,L}\nb_R
+\lampth\ov{\nB}_{1,L}\nb_R
+\hc\Big]\,,
\end{aligned}
\label{eq:lag_f_SU2_components}
\ee
\end{widetext}
where the scalar fields $H$ and $\ns$ still denote the unshifted and unrotated fields defined in Eq.~(\ref{phiDefinition}). 
In order to match with the notation adopted in the previous section, notice that in the unitary gauge
\be
H\equiv
\left(
\begin{array}{c}
0 \\
\nh/\sqrt{2} \\
\end{array}\right)\,.
\ee
\begin{figure}[htb]
\centering
\includegraphics[width=8cm]{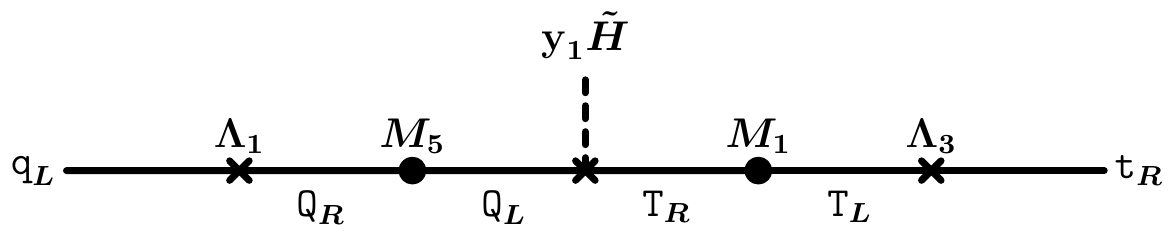} 
\caption{\em Top-quark mass term diagram. }
\label{FigMasses}
\end{figure}

Light fermion masses arise via a series of interactions {\it \`a la} Seesaw and once the Higgs doublets develop a VEV. The diagram 
in Fig.~\ref{FigMasses} exemplifies the top-quark case. A first-order approximation for the values of the top and bottom quark masses is given by
\begin{equation}
m_{t}\approx \, y_1\,\dfrac{\Lambda_1\Lambda_3}{M_1 M_5}\dfrac{v_\nh}{\sqrt{2}}\,,\quad
m_{b}\approx \, y'_1\,\dfrac{\Lambda'_1\Lambda'_3}{M'_1 M'_5}\dfrac{v_\nh}{\sqrt{2}}\,.
\label{eq:leading_contribution_SM_masses}
\end{equation}%

A more precise result can be obtained by diagonalising the whole fermion mass matrix that includes elementary and exotic quarks. It is useful to group all the 
fermions within a single vector, 
\begin{equation}
\Psi=\left(\nK^u,\,\cT,\,\cB,\,\nK'^d\right)^T\,,
\end{equation}
where the ordering of the components is based on their electric charges, $+5/3$, $+2/3$, $-1/3$, and $-4/3$, respectively. Moreover, 
$\cT$ and $\cB$ list together all the states with same electric charge, $+2/3$ and $-1/3$, respectively,
\be
\begin{aligned}
& \cT=\left(\nt,\, \nQ^u,\,\nK^d,\,\nT_5,\,\nT_1,\,\nQ'^{u}\right)^T\,, \\
& \cB=\left(\nb,\,\nQ'^d,\,\nK^u,\,\nB_5,\,\nB_1,\,\nQ^d\right)^T\,.
\end{aligned}
\ee
The whole fermion mass term, still in the interaction basis, can then be written as
\begin{equation}
\sL_\cM=-\ov{\Psi}_L\,\cM(v_\nh,v_\ns)\,\Psi_R,
\end{equation}
where the mass matrix $\cM(v_\nh,v_\ns)$ is a $14\times14$ block diagonal matrix,
\begin{equation}
\cM(v_\nh,v_\ns)=\diag\Big(\mf,\,\cM_\cT(v_\nh,v_\ns),\,\cM_\cB(v_\nh,v_\ns),\,\mpf\Big)\,,\label{eq:BigMassFermion}
\end{equation}
with
\begin{equation}
\cM_\cT(v_\nh,v_\ns)=
\left(
\begin{array}{cccccc}
0 	& \lamo 	& 0 		& 0 		& 0 			& \lampo\\
0 	& \mf 	& 0 		& 0 		& \yo\frac{v_\nh}{\sqrt2} 	& 0\\
0 	& 0		& \mf 	& 0		& \yo\frac{v_\nh}{\sqrt2} 	& 0\\
\lamt	& 0	& 	0	& \mf	& \yo v_\ns		& 0\\
\lamth	& \yt\frac{v_\nh}{\sqrt2}	& \yt\frac{v_\nh}{\sqrt2}	& \yt v_\ns 	& \mo	& 0\\
0	& 0		& 0		& 0		& 0			& \mpf
\end{array}
\right)\,,
\label{eq:FermionMassMatrixComponentsQuarkT}
\end{equation}
and similarly for $\cM_\cB(v_\nh,v_\ns)$, replacing the unprimed parameters with the primed ones and vice-versa. 
The matrix in Eq.~\eqref{eq:BigMassFermion} can be diagonalised through a bi-unitary transformation,
\begin{equation}
\widehat\Psi_{L,R}=U_{L,R}\Psi \quad\Longrightarrow\quad 
\widehat\cM=U_L\,\cM\, U^\dagger_R\,,
\end{equation}
where $\widehat\Psi_{L,R}$ stand for the mass eigenstates and $\widehat\cM$ for the diagonal matrix. The two unitary matrices can be written as block-diagonal structures
\be
U_{L,R}=\diag\left(1,U^\cT_{L,R},U^\cB_{L,R},1\right)\,,
\label{UnitaryMatricesULR}
\ee
where $U^\cT_{L,R}$ and $U^\cB_{L,R}$ diagonalise $\cM_\cT(v_\nh,v_\ns)$ and $\cM_\cB(v_\nh,v_\ns)$, respectively. Finally, the diagonalised mass matrix is given by 
\be
\widehat{\cM}=\diag\left(\mf,\widehat\cM_\cT,\widehat\cM_\cB,\mpf\right)
\ee
and the mass eigenstate fermion fields are defined as 
\be
\begin{aligned}
& \widehat\Psi=\left(K^u,\,\widehat\cT,\,\widehat\cB,\,K'^d\right)^T  \,,  \\
& \widehat\cT=\left( t,\,\sT,\,\sT_2,\,\sT_3,\,\sT_4,\,\sT_5 \right)^T  \,,  \\
& \widehat\cB=\left( b,\,\sB,\,\sB_2,\,\sB_3,\,\sB_4,\,\sB_5 \right)^T \,, 
\label{eq:phys_states}
\end{aligned}
\ee
with both charge $2/3$ and charge $-1/3$ mass eigenstates ordered by increasing masses, so that the lightest states correspond to the top and bottom quarks, 
respectively. The exotic charge states do not mix with the other fields and then $K^u\equiv \nK^u$ and $K^{\prime d}\equiv \nK^{\prime d}$. The full expression for 
the tree level top quark mass is given by
\be
\begin{split}
m_t  =&  \frac{\yo\lamo\lamth v_\nh/\sqrt{2}}{\mo\mf-\yo\yt\left(v^2_h+v^2_\sigma\right)} + \\
&  - \frac{\yo\yt\lamo\lamt v_\ns v_\nh/\sqrt{2}}{\mo\mfsq-\yo\yt \mf\left(v^2_h+v^2_\sigma\right)}\,,
\end{split}
\label{eq:mt}
\ee
and similarly for the bottom quark mass, replacing the unprimed parameters with the primed ones. 
This expression matches the one in Eq.~\eqref{eq:leading_contribution_SM_masses} once the contributions proportional to $y_2$ and $\Lambda_2$ are neglected.

\subsubsection{Interactions with the EW gauge bosons}

A large part of the phenomenological discussion in the next sections will follow from the fermion interactions with the EW gauge bosons, which are modified 
with respect to the SM case due to the mixing between the elementary and the exotic quarks. This is the case for the $Zb\ov{b}$-coupling, the electroweak precision 
observables (EWPOs) and the single production of the VLQs and their decays.

Adopting a compact notation, the Lagrangian terms describing these interactions with the $W^\pm$ and $Z$ gauge bosons read
\be
\begin{aligned}
& \sL_{W}=\frac{g}{\sqrt{2}}\ov{\widehat\Psi}\,\gamma^\mu\left(V_L \ P_L+V_R \ P_R\right) \widehat\Psi \,W^+_\mu+\hc\,,   \\
& \sL_{Z}=\frac{g}{2c_W}\ov{\widehat\Psi}\,\gamma^\mu\left(C_L \ P_L+C_R \ P_R\right) \widehat\Psi \,Z_\mu\,,
\end{aligned}
\label{eq:L_WZ_MLsM}
\ee
where $\widehat\Psi$ are the mass eigenstates in Eq.~\eqref{eq:phys_states}, while the matrices $V$ and $C$ are obtained by means of the unitary matrices 
$U_{L,R}$ in Eq.~\eqref{UnitaryMatricesULR} and the matrices containing the fermion interactions in the interaction basis. Notice that $V_{L,R}$ are 
block off-diagonal and $C_{L,R}$ block diagonal because of charge conservation. $P_{R,L} = (1\pm\gamma_5)/2$ are the usual chirality 
projectors, $g$ is the $SU(2)_L$ gauge coupling and $c_W \equiv \cos \theta_W$, with $\theta_W$ the weak mixing angle.
For $\sL_{W}$, the coupling matrices in this basis read
\begin{equation}
V_{L,R}=U_{L,R} \ \cV_{L,R} \ U^\dagger_{L,R}\,,
\label{VLRdefinition}
\end{equation}
being $\cV_{L,R}$ the $W$-coupling matrices in the interaction basis, whose form can be deduced from Table~\ref{tab:SMTransformations},
\begin{equation}
\cV_{L,R}=
\left(
\begin{array}{cccc}
0 	& \cV^{K^u\mathcal{T}} 	& 0_{1\times6}  & 0 \\
0_{6\times1} 	&  0_{6\times6}	& \cV^{\mathcal{T}\mathcal{B}}_{L,R} & 0_{6\times1}\\
0_{6\times1} 	& 0_{6\times6}	& 0_{6\times6} 	& \cV^{\mathcal{B}K'^d} \\
0	& 0_{1\times6}	& 	0_{1\times6}	& 0 
\end{array}
\right)\,,
\label{eq:V_LR_matrix}
\end{equation}
where
\be
\begin{aligned}
& \cV^{K^u\mathcal{T}}=\left(\cV^{\mathcal{B}K'^d}\right)^\dagger=\Big(0, 0,1,0,0,0\Big) \,, \\
& \cV^{\mathcal{T}\mathcal{B}}_{L}=
\left(
\begin{array}{cccccc}
1 	& 0 	& 0  & 0 & 0 & 0\\
0 	& 0 	& 0 	& 0 & 0 & 1\\
0 	& 0	& 0 	& 0 & 0 & 0\\
0	& 0	& 	0	& 0 & 0 & 0 \\
0	& 0	& 0	& 0 & 0 & 0 \\
0	& 1	& 0	& 0 & 0 & 0
\end{array}
\right) \,,
\end{aligned}
\label{eq:W_L_matrix}
\ee
and $\cV^{\mathcal{T}\mathcal{B}}_{R}$ is identical to  $\cV^{\mathcal{T}\mathcal{B}}_{L}$, except for its $(1,1)$ entry that is vanishing,
\be
\begin{aligned}
& \Big(\cV^{\mathcal{T}\mathcal{B}}_R\Big)_{ij} = \Big(\cV^{\mathcal{T}\mathcal{B}}_L\Big)_{ij} \quad \forall (i,j) \neq (1,1) \,, \\
& \Big(\cV^{\mathcal{T}\mathcal{B}}_{R}\Big)_{1,1}= 0 \,,
\end{aligned}
\label{eq:W_R_matrix}
\ee
corresponding to the fact that right-handed weak eigenstates are $SU(2)_L$ singlets.

For  $\sL_{Z}$, the coupling matrices for mass eigenstates are given by 
\begin{equation}
C_{L,R}= U_{L,R} \ \mathcal{C}_{L,R} \ U^\dagger_{L,R}-2s^2_W\cQ\,,
\label{CLRdefinition}
\end{equation}
with $s_W\equiv \sin \theta_W$, $\cQ$ the electromagnetic charge matrix
\begin{equation}
\cQ=\diag\Big(\dfrac53,\,\dfrac23\, \unity_{6\times6},\,-\dfrac13 \, \unity_{6\times6},\,-\dfrac43\Big)\,,
\end{equation}
and $\mathcal{C}_{L,R}$ the isospin-dependent coupling matrices in the interaction basis
\be
\begin{aligned}
& \mathcal{C}_{L,R}=\diag\Big(1,\,\mathcal{C}^\mathcal{T}_{L,R},\,\mathcal{C}^\mathcal{B}_{L,R},\,1\Big) \,,  \\
& \mathcal{C}^\mathcal{T}_L=-\mathcal{C}^\mathcal{B}_L=\diag\Big(1, 1,-1,0,0,1\Big) \,,  \\
& \mathcal{C}^\mathcal{T}_R=-\mathcal{C}^\mathcal{B}_R=\diag\Big(0, 1,-1,0,0,1\Big) \,.
\end{aligned}
\label{CaligCLRdefinition}
\ee
The normalisation adopted is such that the diagonal entries equal two times the isospin of the weak eigenstates. 

\subsubsection{Interactions with the scalars}

The Lagrangian terms involving the fermion couplings to the scalar fields is
\begin{equation}
\lag_s=\ov{\widehat\Psi}_L\,\text{Y}_h\widehat\Psi_R\, h
+\ov{\widehat\Psi}_L\,\text{Y}_\sigma\widehat\Psi_R\, \s+\hc\,, 
\label{eq:L_s_MLsM}
\end{equation}
where $\text{Y}_h$ and $\text{Y}_\sigma$ are the respective Yukawa coupling matrices of the scalar mass eigenstates, given by
\be
\label{eq:Complete_Yukawa_matrices}
\begin{aligned}
& \text{Y}_h&=U_{L} \ \mathpzc{Y}_h \ U^\dagger_{R}\,, \quad 
\mathpzc{Y}_h&=\diag\left(0,\,\mathpzc{Y}_h^\mathcal{T},\,\mathpzc{Y}_h^\mathcal{B},\,0\right)\,, \\
&\text{Y}_\sigma&=U_{L} \ \mathpzc{Y}_\sigma \ U^\dagger_{R}\,, \quad
\mathpzc{Y}_\sigma&=\diag\left(0,\,\mathpzc{Y}_\sigma^\mathcal{T},\,\mathpzc{Y}_\sigma^\mathcal{B},\,0\right) \,,
\end{aligned}
\ee
with
\be
\begin{aligned}
\mathpzc{Y}_h^\mathcal{T}&=
\left(
\begin{array}{cccccc}
0 	& 0 	& 0  & 0 & 0 & 0\\
0 	& 0 	& 0 	& 0 & -\yo \frac{c_\gamma}{\sqrt{2}} & 0\\
0 	& 0	& 0 	& 0 & -\yo \frac{c_\gamma}{\sqrt{2}} & 0\\
0	& 0	& 	0	& 0 & \yo s_\gamma & 0 \\
0	& -\yt \frac{c_\gamma}{\sqrt{2}}	& -\yt \frac{c_\gamma}{\sqrt{2}}	& \yt s_\gamma & 0 & 0 \\
0	& 0	& 0	& 0 & 0 & 0
\end{array}
\right)\,,  \\
\mathpzc{Y}_h^\mathcal{B}&=\mathpzc{Y}_h^\mathcal{T}\left(\text{y}_i \rightarrow \text{y}'_i\right)\,,  \\ 
\mathpzc{Y}_\sigma^\mathcal{T}&=\mathpzc{Y}_h^\mathcal{T}\left(c_\gamma\rightarrow s_\gamma, \ s_\gamma\rightarrow -c_\gamma \right)\,,  \\
\mathpzc{Y}_\sigma^\mathcal{B}&=\mathpzc{Y}_\sigma^\mathcal{T}\left(\text{y}_i \rightarrow \text{y}'_i\right)\,,
\end{aligned}
\label{eq:H_T_matrix} 
\ee
with $c_\gamma$ and $s_\gamma$ being the cosine and sine of the mixing angle $\gamma$, respectively.

\section{Parameter Space Constraints}
\label{sec:Constraints}

As described in the previous section, the presence of the additional scalar singlet $\s$ and of the exotic fermions modifies the interactions 
of the light fermions with respect to the SM. This section is dedicated to summarise the theoretical and experimental constraints on the parameter space of the \MLsM. 
A few theoretical constraints in the parameter space have already been listed in Section~\ref{SubSect:Scalar}. First of all, $f^2>0$ in order to guarantee 
that $SO(5)\to SO(4)$ breaking takes place. Moreover, $\alpha$ and $\beta$ should satisfy the condition in Eq.~\eqref{TheoContraintsab}. Finally, by 
the diagonalisation of the scalar mass matrix, $\beta>0$ allows a positive mass for the physical $h$ as shown in Eq.~\eqref{TheoContraintsBeta}.

\boldmath
\subsection{Bounds from LHC Higgs searches}\label{sec:scalar_bounds}
\unboldmath

\begin{figure}[h!]
\centering
\includegraphics[width=0.49\textwidth]{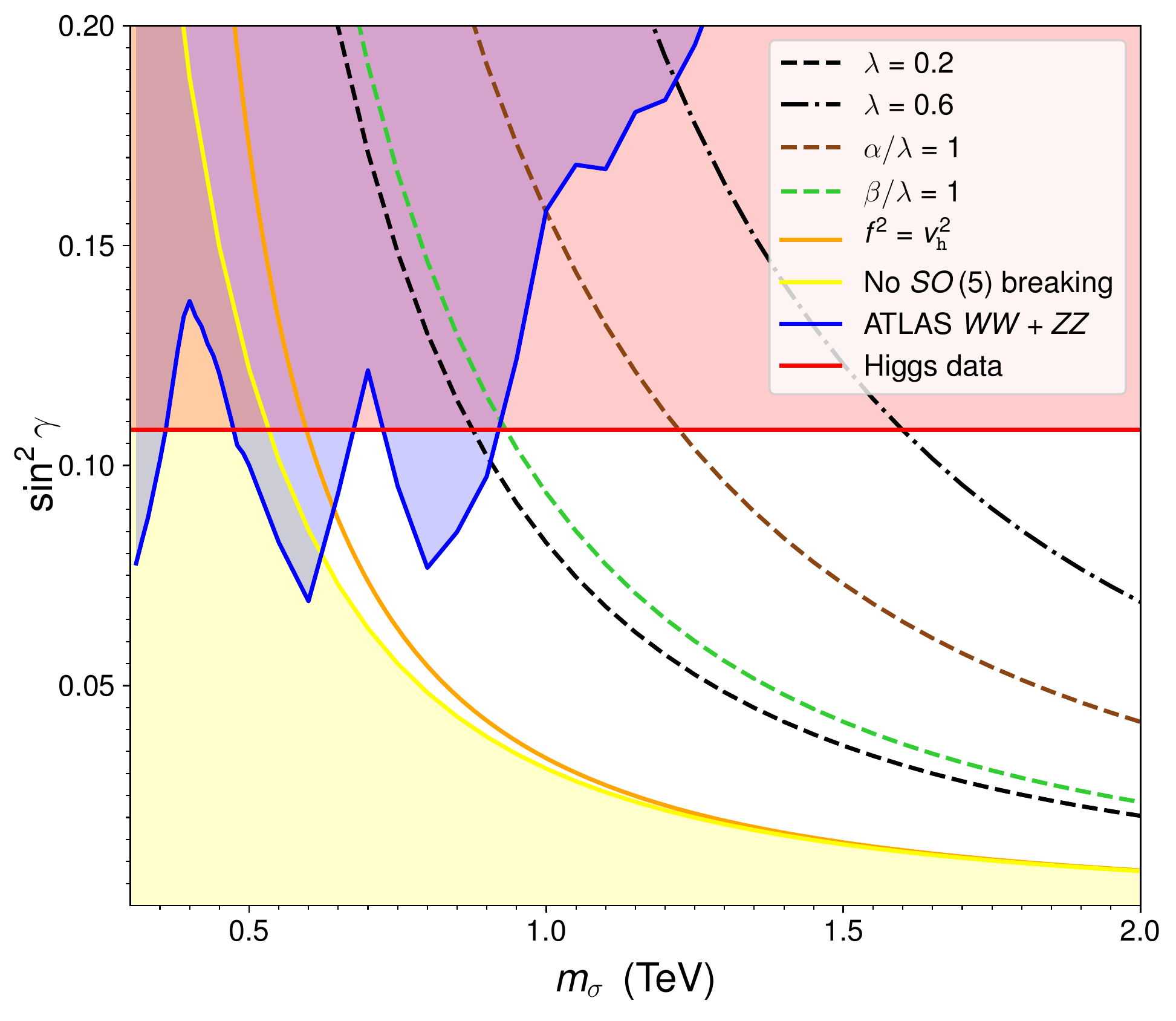} 
\caption{\em Constraints on the mass ($m_\sigma$) and mixing angle ($\sin^2 \gamma$) of the new singlet scalar $\sigma$ (see text for details).}
\label{fig:sg2_VS_ms}
\end{figure}

The measurements of the 125 GeV Higgs signal strengths at the LHC by ATLAS and CMS constrain the Higgs-singlet mixing $\sin \gamma$, which universally suppresses
the couplings of $h$ to SM particles with respect to their SM values.
Very recently, ATLAS has performed a $\sqrt{s} = 13$ TeV analysis of Higgs signal strengths with $80$ fb$^{-1}$ of integrated luminosity~\cite{ATLAS:2019slw}, from which 
we derive the bound 
\be
\sin^2\gamma \lesssim 0.11
\label{sin2gammaBound}
\ee
at 95\% C.L. using a $\chi^2$ fit to the ATLAS data by assuming a universal suppression of Higgs couplings.
This bound is shown in Fig.~\ref{fig:sg2_VS_ms} as an excluded shaded red region in the ($m_\sigma$, $\sin^2 \gamma$) plane.
In addition, Fig.~\ref{fig:sg2_VS_ms} highlights the impact of theoretical constraints on the \MLsM~parameter space: the 
area under the yellow curve is ruled out when requiring a feasible spontaneous symmetry breaking of the \MLsM~$SO(5)$ group. 
Note that realisations where $v_\nh^2/f^2 \equiv \xi = 1$ (with $\xi$ the non-linearity parameter typically introduced in CH models),
depicted by the orange line, and even regions where $v_\nh^2 > f^2$ are not excluded by experimental bounds. 
Finally, the bound from Eq.~\eqref{sin2gammaBound} is consistent with existing bounds 
on $\xi$ in the non-linear limit $m_\sigma\gg m_h$, for which $\xi \simeq \sin^2\gamma$~\cite{Feruglio:2016zvt}.
For completeness, Fig.~\ref{fig:sg2_VS_ms} shows other lines where 
the $SO(5)$ preserving parameter $\lambda$  takes particular values and where it is equal to the remaining parameters of the scalar 
potential of Eq. \eqref{ScalarPotential}.

Additional constraints on the \MLsM~parameter space arise from ATLAS and CMS searches for heavy scalars decaying into SM gauge boson pairs, $WW$ and $ZZ$.
In order to derive the corresponding constraints from these searches, it is useful to introduce
an \textit{effective} Higgs-singlet mixing angle $\sin^2 \gamma_\text{eff}$ 
as the ratio between the cross sections for gluon fusion production of $\sigma$ in the \MLsM \ (taking into account the VLQ loop 
contributions to $gg \to \sigma$, whose expressions can be found in Ref.~\cite{Feruglio:2016zvt}) and for gluon fusion production of a 
SM-like scalar $h_{m_\sigma}$, with mass $m_\sigma$
\begin{equation}
\sin^2 \gamma_\text{eff} = \frac{\sigma(gg \to \sigma)_{\text{ML}\sigma\text{M}}}{\sigma(gg \to h_{m_\sigma})_\text{SM}}\,.
\label{ec:gammaeff}
\end{equation}
In the absence of VLQ loop contributions, one simply has $\sin^2 \gamma_\text{eff} = \sin^2 \gamma$. 
Considering the latest $\sqrt{s} = 13$ TeV ATLAS search for scalar resonances in di-boson final states with 36 fb$^{-1}$ of integrated luminosity~\cite{Aaboud:2018bun}, Fig.~\ref{fig:sg2_VS_ms} shows the the 95\% C.L. limits in $\sin^2 \gamma_\text{eff}$ as an excluded blue area, in the absence of VLQ loop contributions. 
These latter contributions of VLQs to $gg \to \sigma$ are instead included when performing the \MLsM~viable 
parameter space scan in Sec.~\ref{sec:scan}. For the case $\yt=\ypt=0$, these corrections turn out to be 
negligible, and $\sin^2 \gamma_\text{eff} \simeq \sin^2 \gamma$. On the other hand, for $\yt\in\left[3.0,6.0\right]$ (see below) 
the VLQ loop contributions are important
and may interfere constructively with the SM top quark contribution, significantly enhancing the $g g \to \sigma$ production cross section. Numerically the excluded region on the ($m_\sigma$, $\sin^2 \gamma_\text{eff}$) plane corresponds to the blue region in Fig.~\ref{fig:sg2_VS_ms}, replacing $\gamma$ by $\gamma_\text{eff}$, up to small difference due to the variation of the branching ratios of $\sigma\to t\bar t$.

\boldmath
\subsection{Bounds From $Zb\ov{b}$ Coupling}\label{sec:zbb_bounds}
\unboldmath

The couplings of the $Z$ boson to the $c$ and $b$ quarks have been precisely measured at LEP. The tree-level mixing of the bottom quark with the VLQs induces 
deviations from the SM prediction of these couplings (at tree-level for the bottom quark and at loop-level for the charm quark). 
The effective $Zb\ov{b}$ vertex 
can be written as \cite{Aguilar-Saavedra:2013qpa}
\begin{equation}
\sL_{Zbb}=-\frac{g}{2c_W}\ov{b}\,\gamma^\mu\left(c_L \, P_L+c_R \, P_R\right)b\,Z_\mu\,.
\label{eq:zbb}
\end{equation}
The coefficients can be written as the sum of the SM prediction and its deviation,
\begin{equation}
c_{L,R}=c^\text{SM}_{L,R}+\delta c_{L,R}\,,
\label{eq:zbb_shifts}
\end{equation}
where the SM values are given by
\begin{equation}
c^\text{SM}_{L}=1-\frac{2}{3}s^2_W\,,\quad
c^\text{SM}_{R}=-\frac{2}{3}s^2_W\,. 
\label{eq:cLR_SM}
\end{equation}
Comparing Eqs.~\eqref{eq:L_WZ_MLsM} and \eqref{eq:zbb}, the deviations $\delta c_{L,R}$ read
\begin{equation}
\delta c_{L,R}=-\Big(U^{\cB}_{L,R} \ \mathcal{C}^{\cB}_{L,R} \ \left(U^{\cB}_{L,R}\right)^\dagger\Big)_{1,1}-k_{L,R}\,,\label{eq:zbb_MLsM_shifts}
\end{equation}
where $k_L=1$ and $k_R=0$.

Effects of deviations from the SM values can be seen in several observables, such as the ratios of partial 
widths $R_b\equiv\Gamma(b\ov{b})/\Gamma(\text{hadrons})$ and $R_c\equiv\Gamma(c\ov{c})/\Gamma(\text{hadrons})$, the forward-backward 
(FB) charge asymmetry $A^b_\text{FB}$, and the coupling parameter $A_b$ from the LR FB asymmetry~\cite{Bamert:1996px,Aguilar-Saavedra:2013qpa}:
\be
\begin{aligned}
R_b&=R^\text{SM}_b\left(1-1.820\,\delta c_L+0.336\,\delta c_R\right)\,,  \\
A^{b}_\text{FB}&=A^{b,\text{SM}}_\text{FB}\left(1-0.1640\,\delta c_L-0.8877\,\delta c_R\right)\,,   \\
A_{b}&=A^\text{SM}_{b}\left(1-0.1640\,\delta c_L-0.8877\,\delta c_R\right)\,,   \\
R_c&=R^\text{SM}_c\left(1+0.500\,\delta c_L-0.0924\,\delta c_R\right)\,.
\end{aligned}
\ee
The SM values read
\be
\begin{aligned}
R^\text{SM}_b&= 0.21582\,,  \\
A^{b,\text{SM}}_{FB}&= 0.1030\,,  \\
A^\text{SM}_{b}&= 0.9347\,,  \\
R^\text{SM}_c&= 0.17221\,,
\end{aligned}
\ee
while the experimental results that will be used to constrain the parameter space are~\cite{ALEPH:2005ab}
\be
\begin{aligned}
R^\text{exp}_b&=0.21629 \pm 0.00066\,,  \\
A^{b,\text{exp}}_{FB}&=0.0992 \pm 0.0016\,,   \\
A^\text{exp}_{b}&=0.923 \pm 0.020\,,   \\
R^\text{exp}_c&=0.1721 \pm 0.003\,,
\end{aligned}
\label{eq:exp_vals_Zbb_stuff}
\ee
and, with same ordering as in Eq.~\eqref{eq:exp_vals_Zbb_stuff}, the correlation matrix is given by
\begin{equation}
\rho=
\left(
\begin{array}{cccc}
1 	& -0.10 & -0.08  & -0.18 \\
-0.10 	& 1 	& 0.06 	& 0.04 \\
-0.08 	& 0.06	& 1 	& 0.04 \\
-0.18	& 0.04	& 	0.04	& 1
\end{array}
\right) \,.
\label{eq:correl_Zbb_stuff}
\end{equation}

\subsection{Bounds From EWPOs}\label{sec:ewpo_bounds}

The EWPOs also set constraints on the parameter space. In our analysis we restrict ourselves to the $T$ and $S$ oblique parameters, which are the most relevant ones. 
Due to the mixing in the scalar sector, the physical $\s$ acquires couplings with the EW gauge bosons, weighted by $\sin\gamma$, and therefore it 
contributes to the oblique parameters. On the other side, the $h$ couplings get suppressed by $\cos\gamma$ with respect to the SM and this also affects the 
contributions to $T$ and $S$. All in all, the $T$ and $S$ contributions can be straightforwardly derived from the SM one:
\be
\begin{aligned}
T_\text{\MLsM}^h &=c^2_\gamma \, T_\text{SM}^h\,,  \\
T_\text{\MLsM}^\sigma&=s^2_\gamma \ T_\text{SM}^h\left(m_h\rightarrow m_\sigma\right)\,,  \\
S_\text{\MLsM}^h &=c^2_\gamma \, S_\text{SM}^h\,,  \\
S_\text{\MLsM}^\sigma&=s^2_\gamma \ S_\text{SM}^h\left(m_h\rightarrow m_\sigma\right)\,,
\end{aligned}
\ee
so that
\be
\begin{aligned}
\Delta T^{(h, \sigma)} &\equiv  T_\text{\MLsM}^{(h, \sigma)}-T_\text{SM}^h  \\
 &=  s^2_\gamma\left[-T_\text{SM}^h+T_\text{SM}^h\left(m_h\rightarrow m_\sigma\right)\right]\,, \\
\Delta S^{(h,\sigma)} & \equiv S_\text{\MLsM}^{(h , \sigma)}-S_\text{SM}^h \\ 
& =  s^2_\gamma\left[-S_\text{SM}^h+S_\text{SM}^h\left(m_h\rightarrow m_\sigma\right)\right]\,.
\end{aligned}
\label{eq:Scalar_Delta_S}
\ee
The general expression for the $T,S$ parameters considering contributions from VLQs with arbitrary couplings can be found in 
Refs.~\cite{Anastasiou:2009rv,Carena:2006bn}, and are given in appendix~\ref{sec:a} for completeness.
In the SM scenario, that is considering only the SM top and bottom quarks with their EW couplings, the contribution to the $T$ parameter is
\be
\begin{split}
T_\text{SM}^{f} = & \frac{3}{16\pi s^2_W c^2_Wm^2_Z}\times  \\
& \times \left[m^2_t+m^2_b-2\frac{m^2_t m^2_b}{m^2_t-m^2_b}\log\left(\frac{m^2_t}{m^2_b}\right)\right]  \\
\approx & 1.19\,.
\end{split}
\ee
It is then possible to define the deviation from the SM contribution as the difference 
\begin{equation}
\Delta T^f=T^{f}_{\text{ML}\sigma\text{M}}-T_{SM}^{f}\,,
\end{equation}
where $T^{f}_\text{\MLsM}$ corresponds to Eq.~\eqref{eq:T_ferm_generic} considering the whole fermionic content of the \MLsM.
The SM contribution to the $S$ parameter can be obtained considering only the top and bottom contributions in Eq.~\eqref{eq:S_ferm_generic}, and it turns out to be
\begin{equation}
S^{f}_\text{SM}=-\frac{1}{6\pi}\log\left(\frac{m^2_{t}}{m^2_b}\right)\,.
\end{equation}
The deviation from the SM can be expressed in terms of the difference,
\begin{equation}
\Delta S^{f}=S^{f}_\text{\MLsM}-S_\text{SM}^{f}\,,
\end{equation}
where $S^{f}_\text{\MLsM}$ is obtained from Eq.~\eqref{eq:S_ferm_generic} considering the whole fermionic content of the \MLsM.

The sum of scalar and fermionic contributions to both $T$ and $S$ must agree with experimental data,
\be
\begin{aligned}
\Delta T\equiv T -T_{SM}=0.06\pm0.06\, \\
\Delta S\equiv S -S_{SM}=0.02\pm0.07\,
\end{aligned}
\ee
with a correlation of $0.92$.

\section{\MLsM~parameter space scan}\label{sec:scan}

Here we discuss the details of our parameter scan of the \MLsM. For the mass and mixing of the new scalar, two benchmark points 
are chosen within the allowed region in Fig.~\ref{fig:sg2_VS_ms},
\be
\begin{aligned}
& m_{\sigma} = 0.7~\text{TeV} \,,\quad \sin^2 \gamma = 0.1 \,, \\
& m_{\sigma} = 1.0~\text{TeV} \,,\quad \sin^2 \gamma = 0.1 \,.
\end{aligned}
\ee
In the fermionic sector, two different choices of parameters are considered:

\begin{enumerate}

\item Both $\yt$ and $\ypt$ are set to zero, since it is sufficient to have $\yo$ and $\ypo$ different from zero to reproduce the 
measured values of top and bottom quark masses at tree level. With $\yt=\ypt=0$, $\yo$ and $\ypo$ can be determined from 
other parameters and the quark masses in Eq.~(\ref{eq:mt}) which, up to second order, take the form
\be
\begin{split}
m_t = & \yo\frac{v_h}{\sqrt{2}} \frac{\lamo\lamth}{\mo\mf}
\left(1+\frac{\Lambda^2_1}{\mfsq}+\frac{\Lambda'^2_1}{\mpfsq}\right)^{-1/2}\times \\
&\qquad\times \left(1+\frac{\Lambda^2_2}{\mfsq}+\frac{\Lambda^2_3}{\mosq}\right)^{-1/2}\,,
\end{split}
\label{eq:mt_second_order}
\ee
and similarly for the bottom sector, replacing the unprimed parameters with the primed ones and vice-versa. 

\item A Yukawa coupling $\yt$ different from zero and relatively large ($\yt\in\left[3.0,6.0\right]$) is considered. The motivation for this choice is that, with this range of $\yt$, the decay $T \to t \s$ has quite sizeable branching ratio as can be seen below.

\end{enumerate}

For the remaining \MLsM~parameters a numerical scan is performed. After numerically diagonalising 
the mass matrices $\cM_\cT$ and $\cM_\cB$, the constraints discussed in Secs.~\ref{sec:scalar_bounds}--\ref{sec:ewpo_bounds} are enforced, such that all the model predictions for these observables agree with the experimental measurements at $95\%$ C.L.
In addition to the above, the top quark mass resulting from the numerical diagonalisation is required to 
be in the interval $[172,174]$ GeV, and the bottom quark mass in the interval $[4.4,4.8]$ GeV. At the same time, all Yukawa 
couplings are required to be far from the non-perturbativity limit ($\yo$, $\ypo<6$).
The combination of all constraints can be satisfied for $\text{M}_i$, $\text{M}'_i$, $\Lambda_i$ around a few$\TeV$ and $\Lambda'_i$ around a few hundred $\GeV$.

\vspace{1mm}

In the next section, the results of this scan are presented with a focus on the salient phenomenology of the model. In particular, we explore the 
relevant phenomenological features of the lightest exotic quarks, which generally dominate the phenomenology of CH scenarios as discussed previously in this work.

\section{Lightest VLQ Phenomenology}
\label{sec:Phem}

\begin{figure}[t]
\centering
  \begin{minipage}[h]{0.23\textwidth}
    \includegraphics[width=\textwidth]{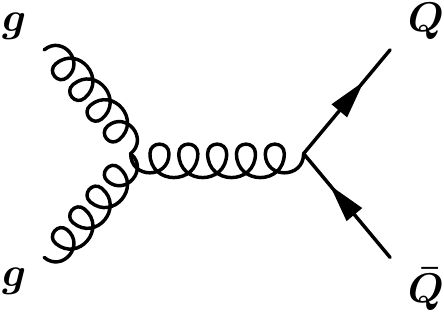}
  \end{minipage}
 \hspace{1mm} 
 \begin{minipage}[h]{0.23\textwidth}
    \includegraphics[width=\textwidth]{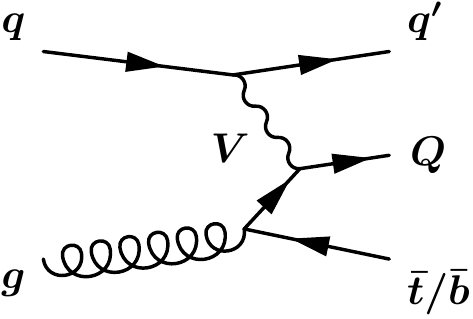}
  \end{minipage}
\caption{\em Feynman diagrams for pair (left) and single (right) production of the VLQ state $Q$. For single VLQ production, $q$ and $q'$ stand for 
generic SM quarks of the first two generations.}
\label{fig:VLQsProduction}
\end{figure}

As it is well-known, VLQs can be produced at colliders like the LHC either in pairs or via single 
production. Examples of Feynman diagrams representing these two possibilities are shown in Fig.~\ref{fig:VLQsProduction}. 
In this work, the focus is on scenarios where the lightest VLQ has electric charge $2/3$ or 
$-1/3$, since it is in this case where the phenomenological features of the \MLsM~may be 
qualitatively different to those of minimal CH models.
Once produced, such lightest VLQ state may decay into a third generation SM quark and a scalar or gauge boson~\cite{Aguilar-Saavedra:2013wba}. 

The Lagrangian terms involved in the production and decay of the lightest exotic quarks of charge $2/3$ ($T$ for the top-partner) and 
$-1/3$ ($B$ for the bottom-partner) read
\be
\begin{aligned}
\sL_{G}  = &g_s\left(\ov{\sT }\gamma^\mu \sT + \ov{\sB}\gamma^\mu \sB \right) \frac{\lambda^a}{2} G^a_{\mu}\,,  \\
\sL_{WQq}  = &\frac{g}{\sqrt{2}}\Big[\ov{\sT}\gamma^\mu\left(V^L_{Tb}\,P_L+V^R_{Tb} \,P_R\right) b+  \\
 & +\ov{t}\gamma^\mu\left(V^L_{tB} \, P_L+V^R_{tB} \, P_R\right) \sB\Big]W^+_\mu+\hc\,,  \\
\sL_{ZQq}  =& \frac{g}{2c_W}\Big[\ov{t}\gamma^\mu\left(X^L_{tT}  \, P_L+X^R_{tT} \, P_R\right) \sT+  \\
 & - \ov{b}\gamma^\mu\left(X^L_{bB} \, P_L+X^R_{bB} \, P_R\right) \sB\Big] Z_\mu +\hc\,, \\
\sL_{hQq}  =& \Big[\ov{t}\left(y^{h,L}_{tT}  \, P_L+y^{h,R}_{tT} \, P_R\right) \sT+  \\
 & + \ov{b}\left(y^{h,L}_{bB} \, P_L+y^{h,R}_{bB} \, P_R\right) \sB\Big] h +\hc\,, \\
\sL_{\s Qq}  =& \Big[\ov{t}\left(y^{\s,L}_{tT}  \, P_L+y^{\s,R}_{tT} \, P_R\right) \sT+  \\
 & + \ov{b}\left(y^{\s,L}_{bB} \, P_L+y^{\s,R}_{bB} \, P_R\right) \sB\Big] \s +\hc\,,
\label{eq:LVLQ_qSM}
\end{aligned}
\ee
where $\lambda^a$ are the Gell-Mann matrices, $a$ is a colour index, $G^a_{\mu}$ are the gluon fields
and $g_s$ is the $SU(3)_c$ gauge coupling. In terms of the $V$ 
matrices introduced in Eq.~\eqref{VLRdefinition},
\be
\begin{aligned}
V^{L,R}_{Tb}= \Big(V_{L,R}\Big)_{3,8} =\Big(U^{\cT}_{L,R} \ \cV^{\cT\cB}_{L,R} \ \left(U^{\cB}_{L,R}\right)^\dagger\Big)_{2,1} \\
V^{L,R}_{tB}= \Big(V_{L,R}\Big)_{2,9} = \Big(U^{\cT}_{L,R} \ \cV^{\cT\cB}_{L,R} \ \left(U^{\cB}_{L,R}\right)^\dagger\Big)_{1,2},
\end{aligned}
\label{eq:V_coeffs_MLsM}
\ee
being $U^{\cT,\cB}_{L,R}$ the rotation matrices given at Eq.~\eqref{UnitaryMatricesULR} and $\cV^{\cT\cB}_{L,R}$ the matrices defined 
at Eqs.~\eqref{eq:W_L_matrix} and~\eqref{eq:W_R_matrix}. In terms of the $C$ matrices given at Eq.~\eqref{CLRdefinition},
\be
\begin{aligned}
X^{L,R}_{tT}&= \Big(C_{L,R}\Big)_{2,3} =\Big(U^{\cT}_{L,R} \ \mathcal{C}^{\cT}_{L,R} \ \left(U^{\cT}_{L,R}\right)^\dagger\Big)_{1,2}  \\
-X^{L,R}_{bB}&= \Big(C_{L,R}\Big)_{8,9}=\Big(U^{\cB}_{L,R} \ \mathcal{C}^{\cB}_{L,R} \ \left(U^{\cB}_{L,R}\right)^\dagger\Big)_{1,2}\,,
\end{aligned}
\label{eq:X_coeffs_MLsM}
\ee
where $\mathcal{C}^{\cT,\cB}_{L,R}$ are defined at Eqs.~\eqref{CaligCLRdefinition}.
Finally, the Yukawa couplings of Eq.~\eqref{eq:LVLQ_qSM} are given by
\be
\begin{aligned}
y^{s,L}_{tT}=\Big(\text{Y}_s\Big)_{3,2}=\Big(U^{\cT}_{L}\mathpzc{Y}_s^\cT\left(U^{\cT}_{R}\right)^\dagger\Big)_{2,1} \\
y^{s,R}_{tT}=\Big(\text{Y}_s\Big)_{2,3}=\Big(U^{\cT}_{L}\mathpzc{Y}_s^\cT\left(U^{\cT}_{R}\right)^\dagger\Big)_{1,2} \\
y^{s,L}_{bB}=\Big(\text{Y}_s\Big)_{9,8}=\Big(U^{\cB}_{L}\mathpzc{Y}_s^\cB\left(U^{\cB}_{R}\right)^\dagger\Big)_{2,1} \\
y^{s,R}_{bB}=\Big(\text{Y}_s\Big)_{8,9}=\Big(U^{\cB}_{L}\mathpzc{Y}_s^\cB\left(U^{\cB}_{R}\right)^\dagger\Big)_{1,2},
\end{aligned}
\ee
with $s$ being either $h$ or $\sigma$, $\text{Y}_s$ given in Eq.~\eqref{eq:Complete_Yukawa_matrices} and $\mathpzc{Y}_s^{\cT,\cB}$ in Eq.~\eqref{eq:H_T_matrix}.

\vspace{1mm}

In the following we consider separately the scenarios where the lightest VLQ has charge $-1/3$ ($B$) 
or charge $2/3$ ($T$).

\boldmath
\subsection{Bottom partner ($B$) phenomenology}
\unboldmath

\begin{figure}[h!]
\centering
\includegraphics[width=0.49\textwidth]{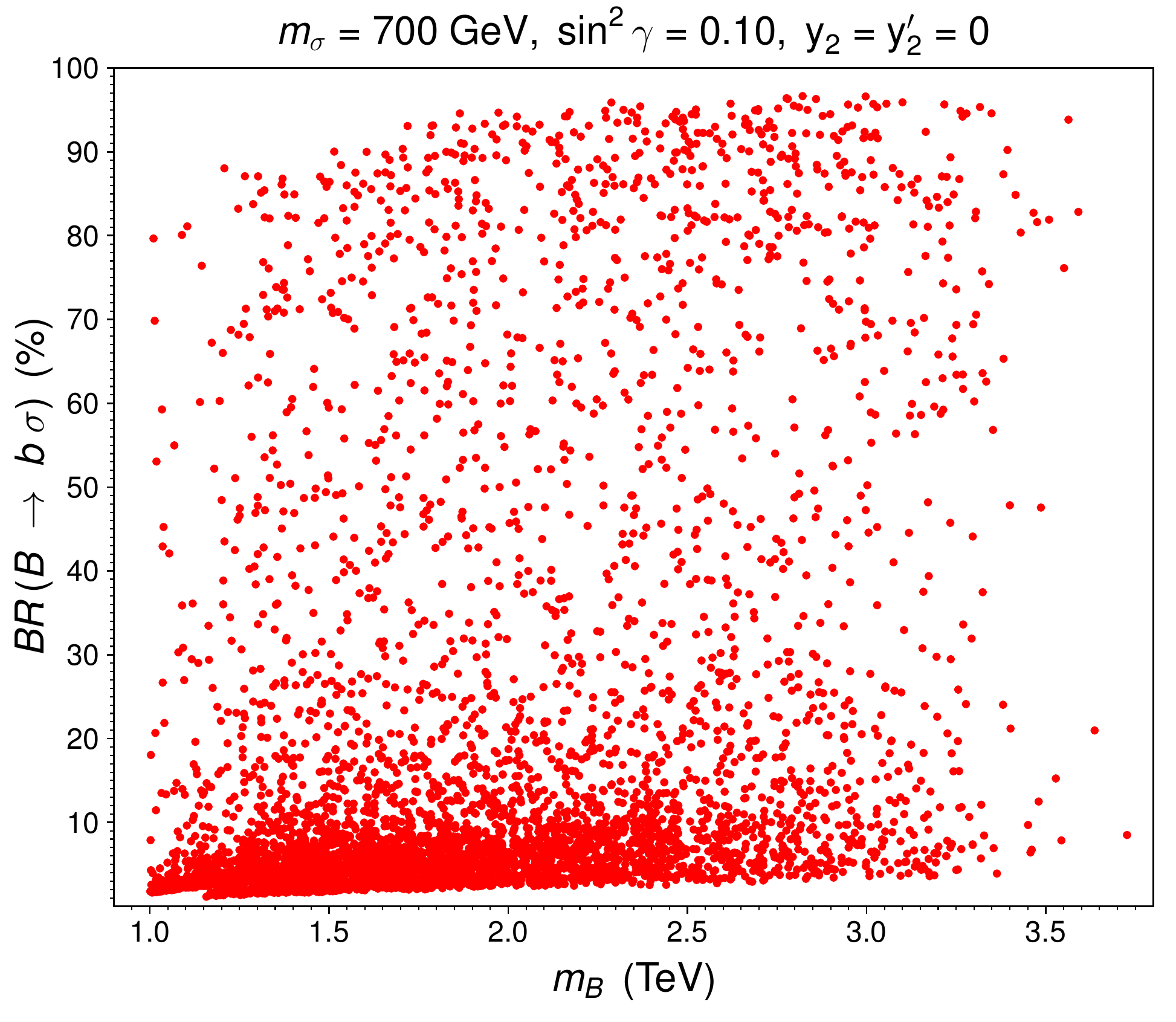} 
\caption{\em Values of the branching ratio $BR(B \to \s b)$ as a function of the bottom partner mass $m_B$ for the allowed points from our \MLsM~parameter scan 
(only points where $B$ is the lightest VLQ are included).}
\label{fig:all_y2_zero_BR_BP}
\end{figure}

For scenarios with the $B$ state as lightest VLQ the most salient phenomenological feature is the large branching 
ratio $BR(B \to \sigma b)$ that is possible in wide regions of the parameter space. This non-standard VLQ decay may even be the dominant one. 
In Fig.~\ref{fig:all_y2_zero_BR_BP}, the results of our \MLsM~scan for $m_{\sigma} = 700$ GeV and $\yt=\ypt=0$ are shown 
for $BR(B \to \sigma b)$ as a function of the bottom partner mass $m_{B}$ (the varying density of points has no physical meaning but is rather an artifice of 
the parameterisation used in the scan), showing that values $BR(B \to \sigma b) > 0.5$ may easily be obtained. The corresponding results for 
$m_{\sigma} = 1$ TeV are found to be quantitatively very similar.
A specific example of model parameters that result in the lightest VLQ being a bottom partner state $B$ predominantly decaying into $\sigma b$ is given 
in the first line of Tab.~\ref{tab:three_benchmarks}.

\begin{table*}[htb]
\centering 
\renewcommand{\arraystretch}{1.5}
\footnotesize
\begin{tabular}{|c|c|c|c|c|c|c|c|c|c|c|c|c|c|c|c|c|c|c|c|c|c|c|c|c|c|} 
\hline
$m_\sigma$ & $\mo$ & $\mf$ & $\mpo$ & $\mpf$ & $\lamo$ & $\lamt$ & $\lamth$ & $\lampo$ & $\lampt$ & $\lampth$ & $\yo$ & $\yt$ & $\ypo$ & $m_T$ & $m_B$ & $V^{L}_{Tb}$ & $V^{R}_{Tb}$ & $X^{L}_{bB}$ & $X^{R}_{bB}$ & $BR$ & $BR$\\
$\left(\text{TeV}\right)$ & $\left(\text{TeV}\right)$ & $\left(\text{TeV}\right)$ & $\left(\text{TeV}\right)$ & $\left(\text{TeV}\right)$ & $\left(\text{TeV}\right)$ & $\left(\text{TeV}\right)$ & $\left(\text{TeV}\right)$ & $\left(\text{TeV}\right)$ & $\left(\text{TeV}\right)$ & $\left(\text{TeV}\right)$ & & & & $\left(\text{TeV}\right)$ & $\left(\text{TeV}\right)$ & & & & & $\left(\s t\right)$ & $\left(\s b\right)$ \\
\hline\hline
$0.7$ & $1.2$ & $6.2$ & $2.4$ & $1.6$ & $5.0$ & $1.2$ & $1.8$ & $1.0$ & $0.1$ & $0.7$ & $2.1$ & $0$ & $0.2$ & $1.8$ & $1.6$ & $0$ & $0.005$ & $0.001$ & $0.003$ & $0.04$ & $0.84$\\
\hline
$0.7$ & $1.0$ & $6.3$ & $4.6$ & $3.6$ & $7.7$ & $3.1$ & $3.3$ & $0.7$ & $0.6$ & $0.9$ & $1.4$ & $4.6$ & $1.2$ & $1.7$ & $3.5$ & $0.030$ & $0$ & $0.002$ & $0.001$ & $0.57$ & $0.82$ \\
\hline
$1.0$ & $1.7$ & $6.1$ & $3.6$ & $3.2$ & $8.9$ & $1.9$ & $3.1$ & $0.8$ & $0.4$ & $0.6$ & $1.4$ & $4.9$ & $1.1$ & $1.9$ & $3.0$ & $0.045$ & $0$ & $0.004$ & $0$ & $0.29$ & $0.56$ \\
\hline 
\end{tabular}
\caption{\em Three different benchmark points as examples. In all cases $\sin^2\gamma=0.1$ and $\ypt=0$, as explained above. 
The first two lines refer to the $m_\sigma = 700$ GeV Benchmark, while the third line to the $m_\sigma = 1$ TeV Benchmark. 
The example in the first line illustrate the case with a lightest $B$ quark, while the other two examples the case in which the $T$ quark is the lightest VLQ.}
\label{tab:three_benchmarks}
\end{table*}

\begin{figure}[h!]
\centering
\includegraphics[width=0.49\textwidth]{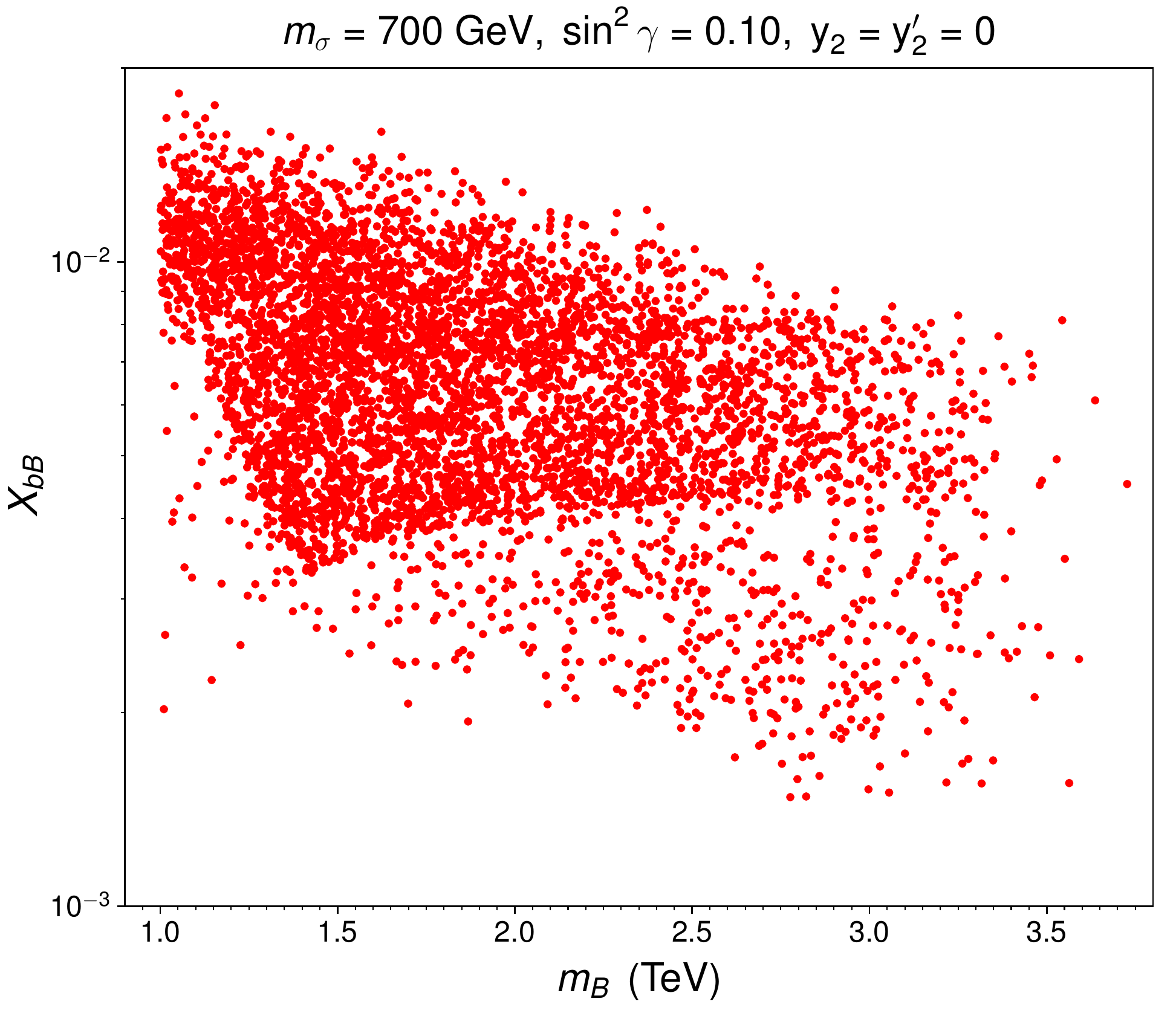}
\caption{\em Values of the $X_{bB}$ coupling in Eq.~(\ref{ec:Xbb}) as a function of the bottom partner mass $m_B$ for the allowed points 
from our \MLsM~parameter scan (only points where $B$ is the lightest VLQ are included).}
\label{fig:all_y2_zero_X_sqrt}
\end{figure}

\begin{figure}[h!]
\centering
\includegraphics[width=0.49\textwidth]{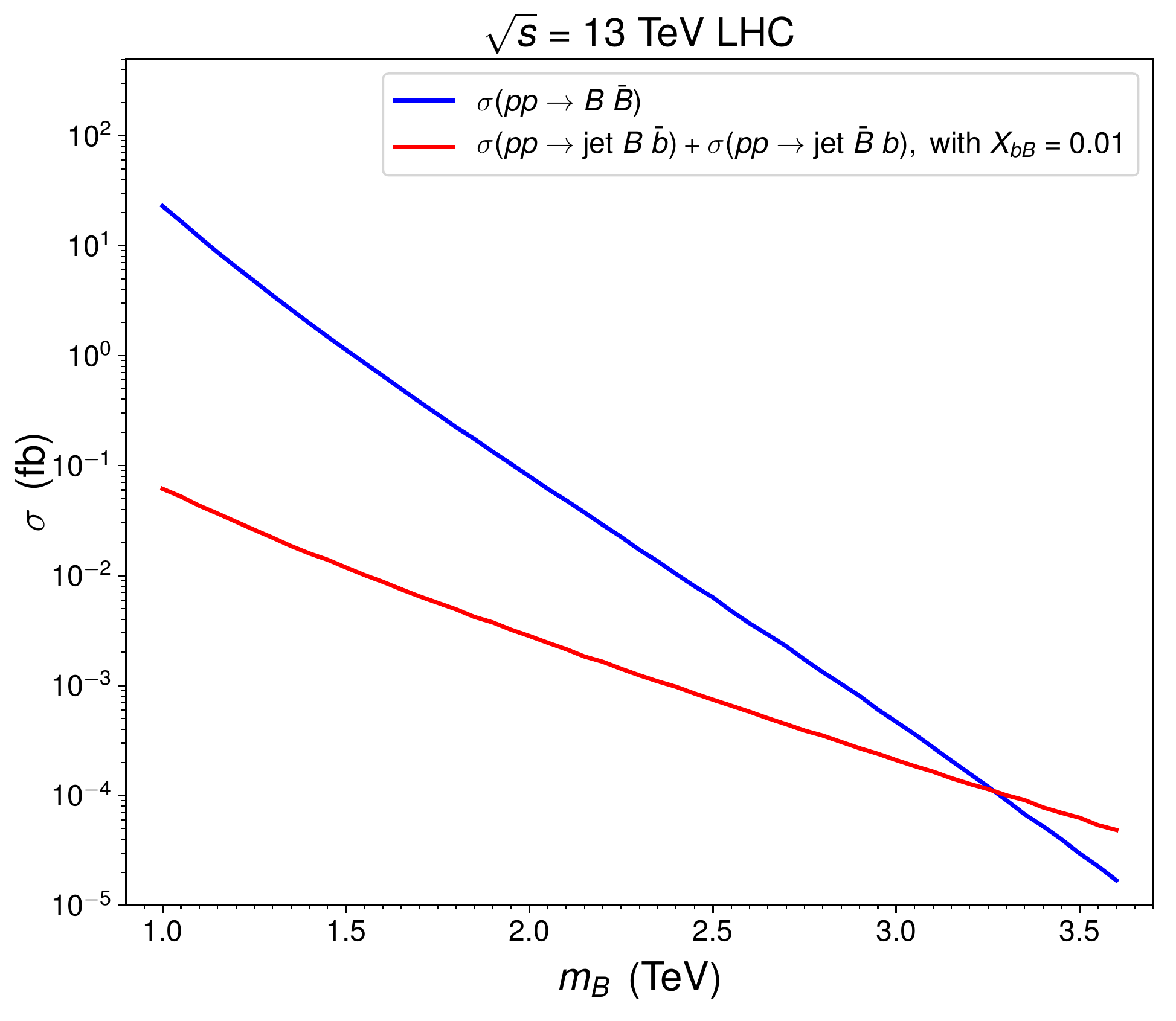}
\caption{\em Cross sections  for $B$ pair production (blue) and single production with $X_{bB} = 0.01$ (red) at 13 TeV LHC, as a function of the VLQ mass $m_B$.}
\label{fig:production_B}
\end{figure}

The dominant contribution to the single $B$ production cross section is given by the processes\footnote{Other single $B$ production channels 
like $p p \to B \bar{t} q$ and $p p \to \bar{B} t q$, proportional to $V_{tB}^2 = \left(V_{tB}^L\right)^2 + \left(V_{tB}^R\right)^2$, yield a 
subdominant contribution.} 
$p p \to B \bar{b} q$, $p p \to \bar{B} b q$ 
(through the Feynman diagram in Fig.~\ref{fig:VLQsProduction} (right), with $V$ being the $Z$ boson), which are controlled by 
the strength of the $BbZ$ flavour-changing neutral coupling defined in terms of the 
couplings from Eq.~\eqref{eq:LVLQ_qSM} as
\begin{equation}
X_{bB} = \sqrt{\left(X_{bB}^L\right)^2 + \left(X_{bB}^R\right)^2 }\,.
\label{ec:Xbb}
\end{equation}
Fig.~\ref{fig:all_y2_zero_X_sqrt} shows the size of $X_{bB}$ as a function of $m_B$ for 
our \MLsM~scan parameter points with $m_{\sigma} = 700$ GeV and $\yt=\ypt=0$. The values for 
$X_{bB}$ barely exceed $0.02$ as seen in Fig.~\ref{fig:all_y2_zero_X_sqrt}, 
resulting in single production cross sections for $B$ much smaller than the 
pair production cross section $\sigma(p p \to B \bar{B})$ at $\sqrt{s} = 13$ TeV, except for very large mass
values $m_B \gtrsim 3.5$ TeV. This can be seen explicitly in
Fig.~\ref{fig:production_B}, which 
shows the $B$ pair production cross section 
(independent of $X_{bB}$) and the single $B$ production for $X_{bB} = 0.01$, 
both obtained with {\tt Madgraph$\_$aMC@NLO}~\cite{Alwall:2014hca}, using a {\tt Feynrules}~\cite{Alloul:2013bka} implementation of the model in 
Eq.~\eqref{eq:LVLQ_qSM}.

\boldmath
\subsection{Top partner ($T$) phenomenology}
\unboldmath

\begin{figure}[h!]
\centering
\includegraphics[width=0.49\textwidth]{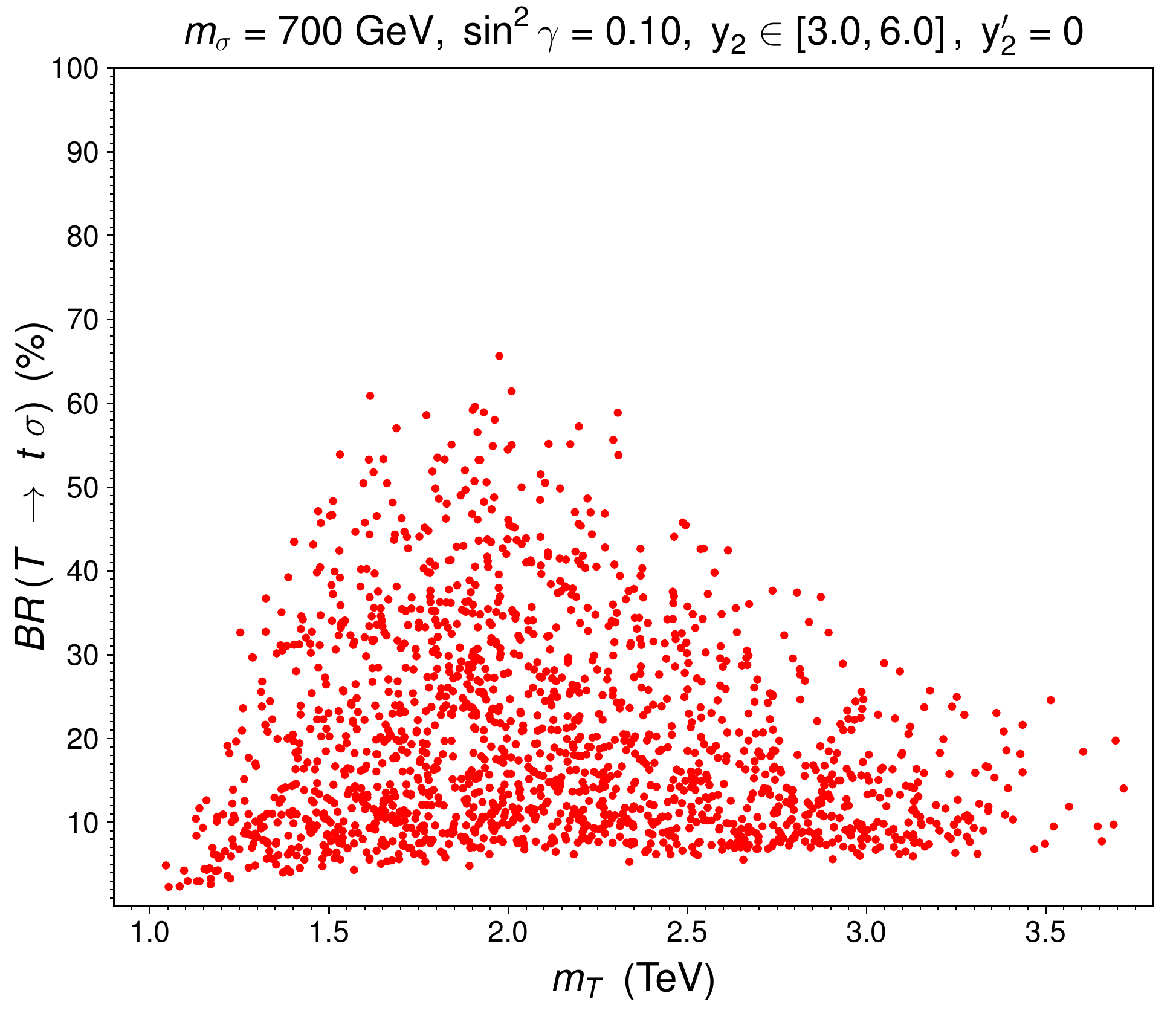} 
\includegraphics[width=0.49\textwidth]{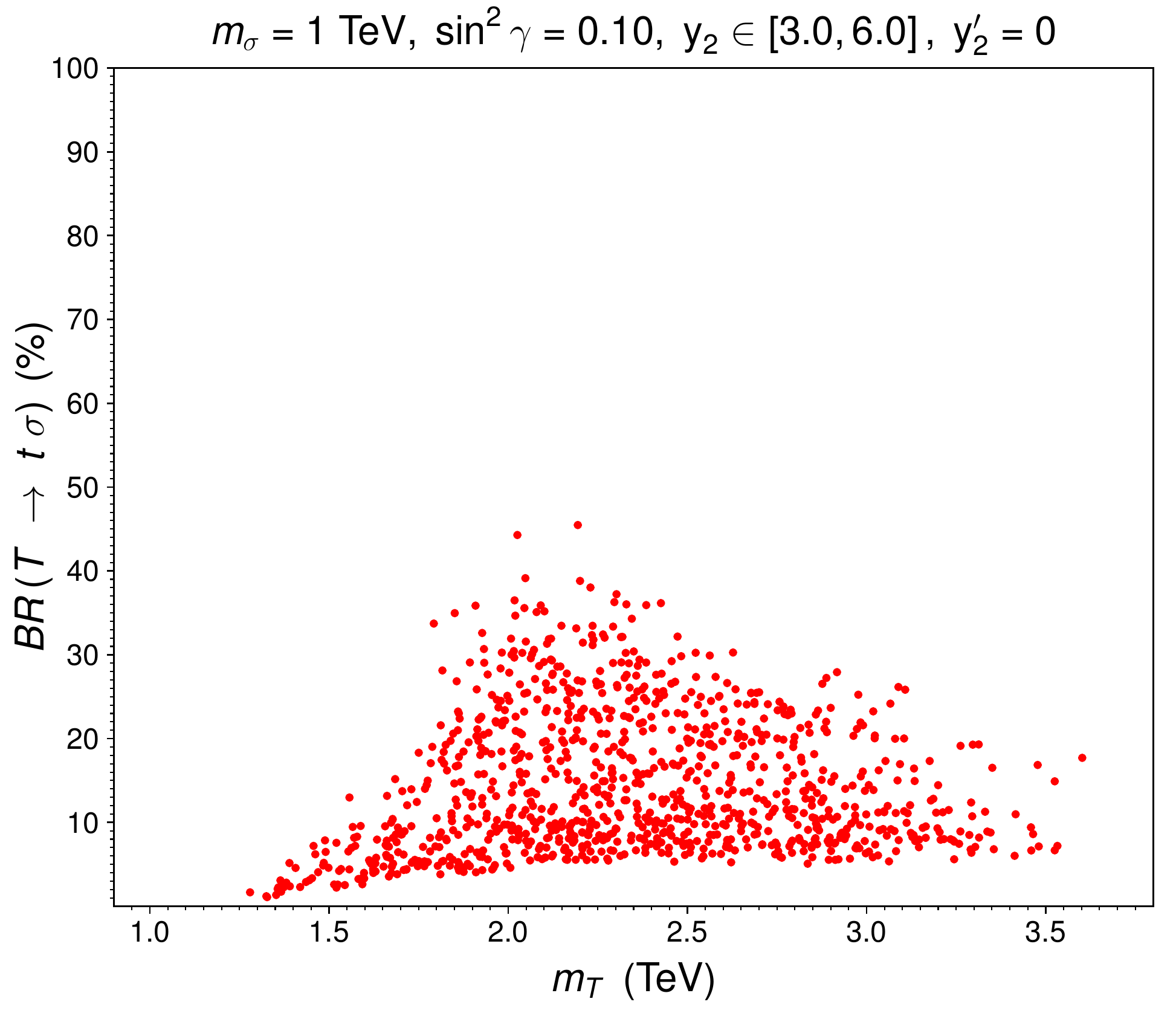}
\caption{\em Values of the branching ratio $BR(T \to \s t)$ as a function of the top partner mass $m_T$ and $m_\sigma = 700$ GeV (upper panel), 
$m_\sigma = 1$ TeV (lower panel) for the allowed points from our \MLsM~parameter scan 
(only points where $T$ is the lightest VLQ are included).}
\label{fig:y2_BIG_yp2_zero_BR_T_vs_MT}
\end{figure}

For scenarios with the $T$ state as lightest VLQ, the branching ratio for the non-standard  decay $T \to \sigma t$ is generally small for $\yt=\ypt=0$
but can be quite large for sizable $\yt$, as shown 
in Fig.~\ref{fig:y2_BIG_yp2_zero_BR_T_vs_MT} respectively for $m_\sigma = 700$ GeV (upper panel) and $m_\sigma = 1$ TeV (lower panel).
The increase in $m_\sigma$ generically leads however to a decrease in the maximum possible size for $BR(T \to \sigma t)$, as seen by comparing 
the upper and lower panels of Fig.~\ref{fig:y2_BIG_yp2_zero_BR_T_vs_MT}.
The second and third lines of Tab.~\ref{tab:three_benchmarks} respectively give an example of 
model parameters that result in the lightest VLQ being a $T$ quark with a large branching ratio to $\sigma t$ for $m_\sigma = 700$ GeV
and $m_\sigma = 1$ TeV.

\begin{figure}[h!]
\centering
\includegraphics[width=0.49\textwidth]{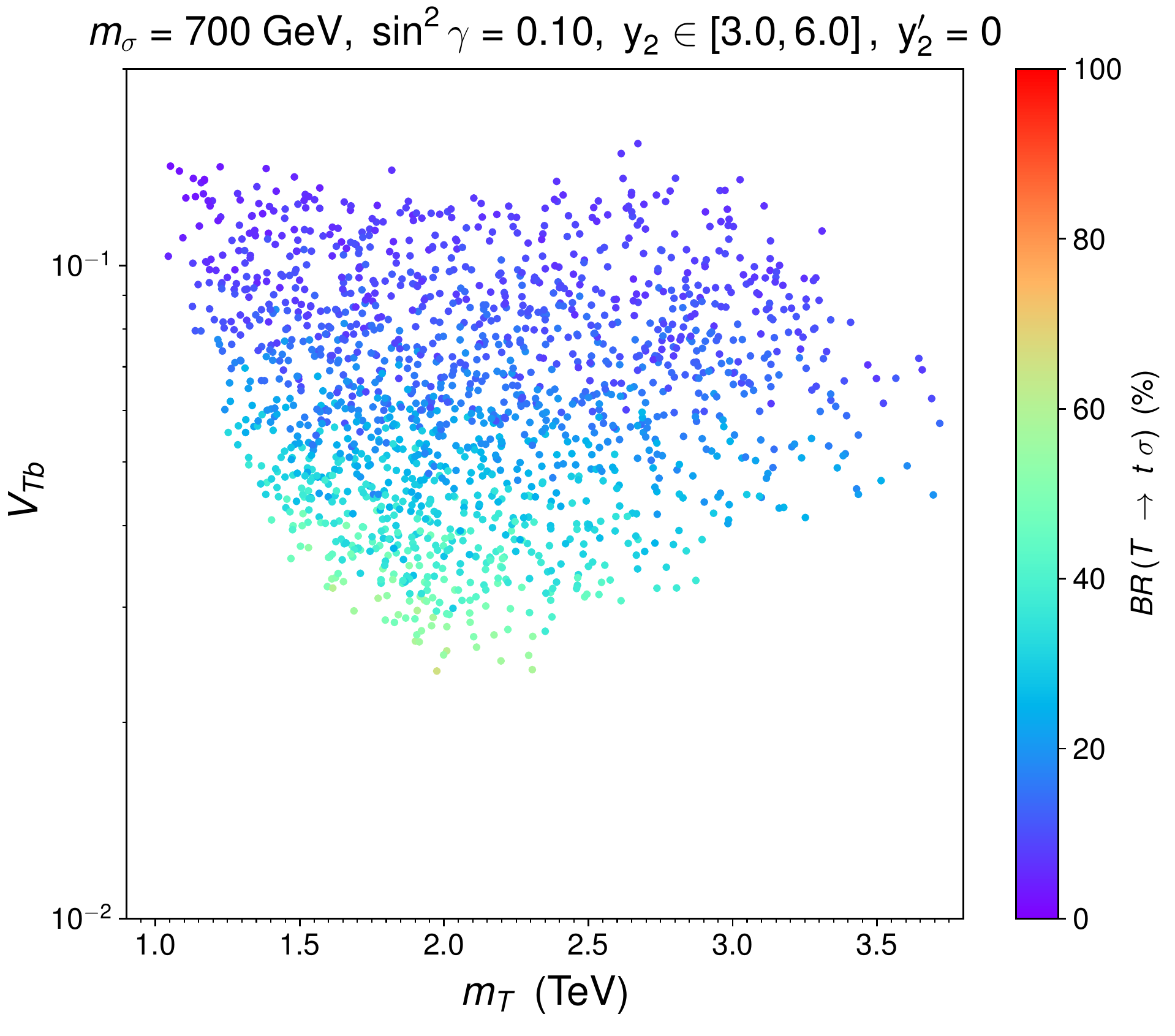}
\includegraphics[width=0.49\textwidth]{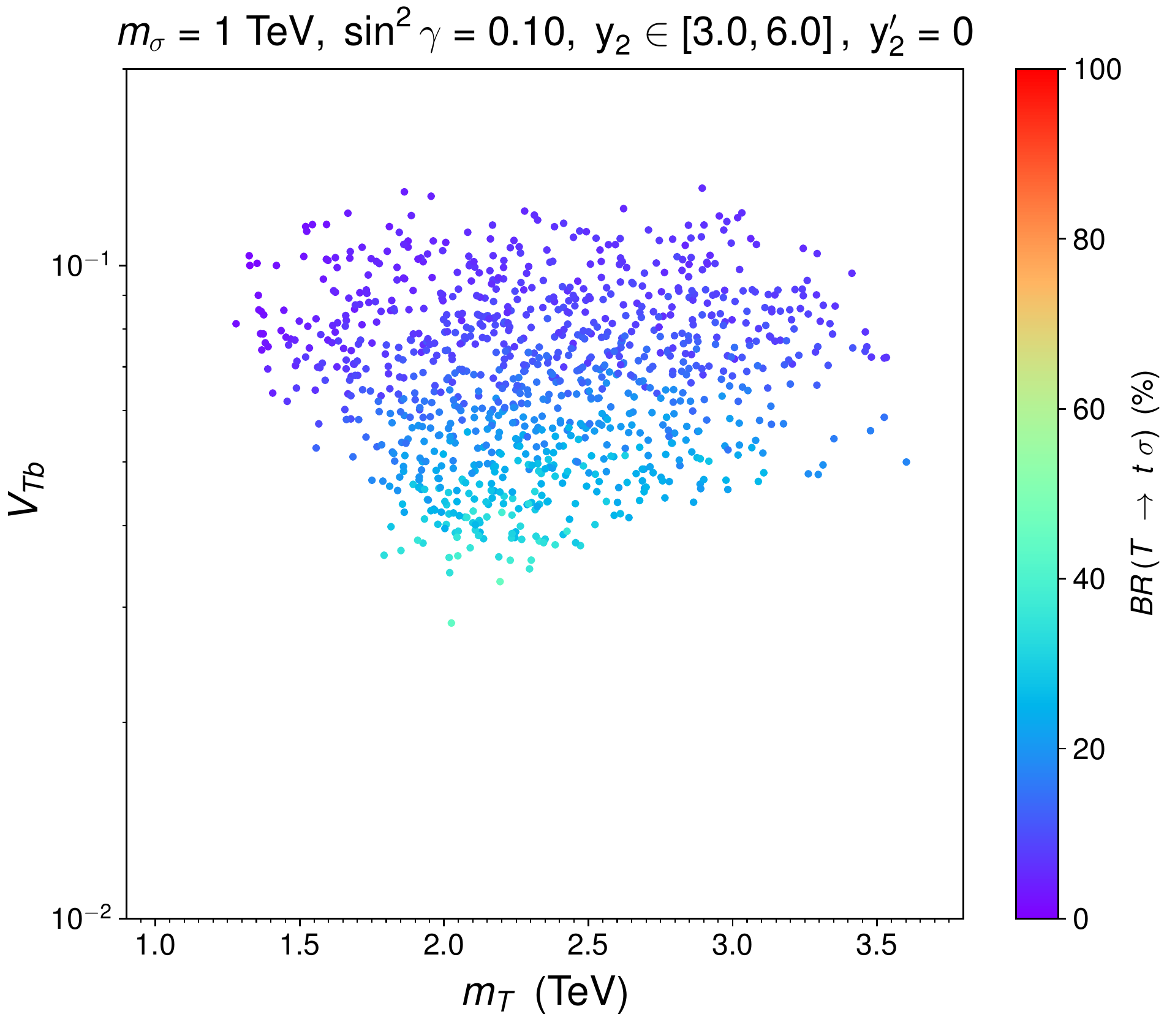}
\caption{\em Values of the $V_{Tb}$ coupling in Eq.~(\ref{ec:VTb}) as a function of the top partner mass $m_T$, for $\yt\in\left[3.0,6.0\right]$ and 
respectively $m_\sigma = 700$ GeV (upper panel) and $m_\sigma = 1$ TeV (upper panel), for the allowed points 
from our \MLsM~parameter scan (only points where $B$ is the lightest VLQ are included). The color of the scan points corresponds to its range of 
values for $BR(T \to \s t)$ (see text for details).}
\label{fig:y2_BIG_yp2_zero_V_Left}
\end{figure}

The mixing of the $T$ state with the SM top and bottom quarks can be sizeable in these scenarios, leading to large values of the single 
$T$ production cross section, but this is anti-correlated with the presence of a large 
$BR(T \to \sigma t)$: for large mixing the branching ratio to $\sigma t$ is smaller, and vice versa. 
Fig.~\ref{fig:y2_BIG_yp2_zero_V_Left} shows the allowed values for the 
coupling $V_{Tb}$, given in terms of the couplings from Eq.~\eqref{eq:LVLQ_qSM} as 
\begin{equation}
V_{Tb} = \sqrt{\left(V_{Tb}^L\right)^2 + \left(V_{Tb}^R\right)^2 }
\label{ec:VTb}
\end{equation}
as a function of the $T$ mass $m_T$. The different colours of the scan points 
correspond to different ranges of $BR(T \to t \sigma)$. As Fig.~\ref{fig:y2_BIG_yp2_zero_V_Left} highlights, sizeable mixings 
$V_{Tb} \simeq 0.1$ are possible, but only for a small 
branching ratio for $T \to t \sigma$. Notice that there appears to be a {\em minimum} value of $V_{Tb}$ as a function of $m_T$: 
$V_{Tb}$ can indeed be below such apparent minima in Fig.~\ref{fig:y2_BIG_yp2_zero_V_Left}, but in such 
case a charge $-1/3$ $B$ state is lighter than $T$.

\begin{figure}[h!]
\centering
\includegraphics[width=9cm]{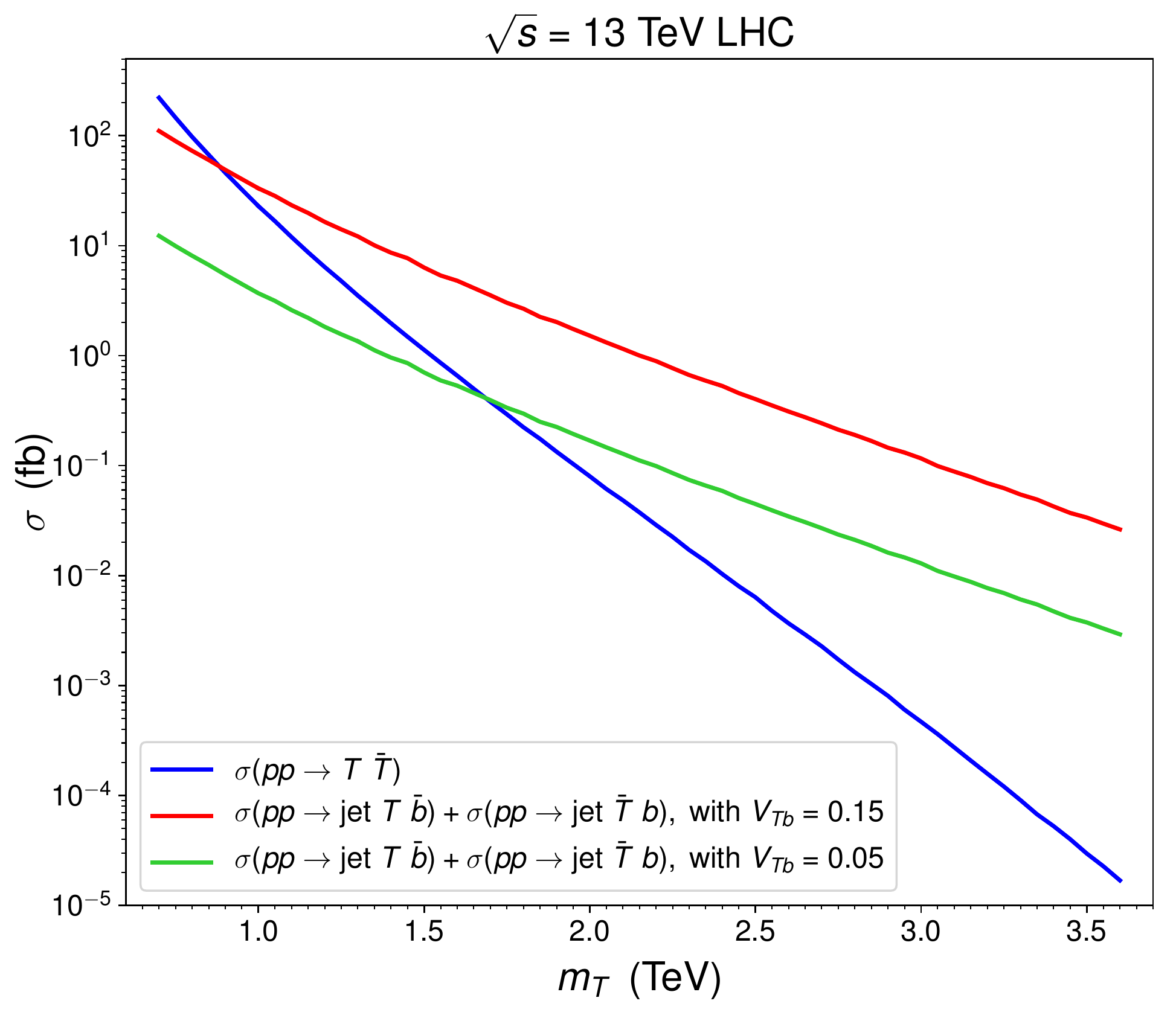}
\caption{\em $T$ pair production (blue) and single production for $V_{Tb} = 0.15$ (red) and $V_{Tb} = 0.05$ (green) cross sections at 13 TeV LHC, 
as a function of the VLQ mass $m_T$.}
\label{fig:production_T}
\end{figure}

The coupling $V_{Tb}$ controls the dominant contributions to single $T$ production via the processes
$p p \to T \bar{b} q$, $p p \to \bar{T} b q$ through the Feynman diagram in Fig.~\ref{fig:VLQsProduction} (right), with $V$ being the $W$ boson. 
Figure~\ref{fig:production_T} shows together the $T$ pair production (independent of $V_{Tb}$) and the single $T$ production cross sections for two benchmark 
values $V_{Tb} = 0.15, 0.05$, computed with {\tt Madgraph$\_$aMC@NLO}~\cite{Alwall:2014hca}.
It becomes clear that for $\mathcal{O}(0.1)$ mixings $V_{Tb}^L$, the single $T$ production cross section becomes more important than the pair production one 
above $m_T \simeq 1$ TeV.

Finally, let us comment that a general feature of models with more than just one VLQ state or multiplet is that the mixing of 
charge-$2/3$ exotic quarks $T$ with the third generation SM quarks can be larger than in minimal models~\cite{Aguilar-Saavedra:2013qpa}. As shown, this also 
happens in the \MLsM~due to the interplay of different contributions to EWPO from the 
several new fermions and the new scalar. From the experimental point of view, the possibility of large mixing is very interesting, as 
it leads to much larger single $T$ production cross sections, which can in fact dominate the overall production of VLQ states above $m_Q \simeq 1$ TeV.

\boldmath
\subsection{Current LHC sensitivity to $B \to \sigma b$ and $T \to \sigma t$}
\unboldmath

As highlighted in the above sections, a salient characteristic of the \MLsM~is the possibility of exotic decays of the new VLQ states, 
$B \to \sigma b$ and $T \to \sigma t$. The singlet state $\sigma$ dominantly decays into a pair of SM gauge bosons or a pair of 125 GeV Higgses, the 
branching fraction into other SM states (e.g.~$\sigma \to t \bar{t}$) being generally below 10\%. 
The exotic decays of the vector-like $T$, $B$ quarks via Eqs.~\eqref{ec:TBexo} and~\eqref{ec:TBexo2} then produce additional $W$, $Z$ or Higgs bosons 
compared to the standard VLQ decays in Eq.~\eqref{ec:TBstd}. 
In a resolved regime in which the $\sigma$ state is not very boosted (as a result of $m_{\sigma}$ and $m_{T/B}$ not being very far apart, 
which is something to be expected in the \MLsM), its decay products are 
well separated and the signals resulting from single or pair production of VLQs followed by the exotic decays are similar to the 
ones already searched for at the LHC (i.e. the standard decays), but with higher SM gauge/Higgs boson multiplicities. 
It is then pertinent to ask ourselves to which extent the existing ATLAS and CMS searches are sensitive to 
the $B \to \sigma b$ and $T \to \sigma t$ decay signatures. The answer depends, case by case, on how {\it inclusive} the event selection for 
these experimental searches is. Several examples of LHC searches for VLQs that are sensitive to the new decay modes introduced in this work 
are:

\vspace{2mm}

\noindent 1. {\em ATLAS search for same-sign di-leptons or three leptons plus $b$-tagged jets~\cite{Aaboud:2018xpj}.}
This analysis primarily targets the process 
\be
\begin{aligned}
pp  \to B \bar B  \to  W^- t \, W^+ \bar t \to  W^- W^+ b \, W^+ W^- \bar b \,,
\end{aligned}
\label{ec:BBWt}
\ee
with two same-sign or three leptons produced from the decays of the $W$ bosons. The events are required to have a large 
amount of missing transverse energy $\etmiss$, as well as large transverse energy $H_T$ (defined as the scalar sum of all transverse momenta of jets and leptons). 
This search is also sensitive to the processes $p p \to T \bar T \to \sigma t \,\sigma \bar t$ and $p p \to B \bar B \to \sigma b \, \sigma \bar b$, 
with $\sigma \to W^+ W^-$. In the former case, the final state contains two extra $W$ bosons compared to Eq.~\eqref{ec:BBWt}, 
and hence the combinatorial factor to get two same-sign or three leptons is larger. 
The latter process yields the same final state as Eq.~\eqref{ec:BBWt}. 

\begin{figure}[h!]
\centering
\begin{tabular}{c}
\includegraphics[width=0.488\textwidth]{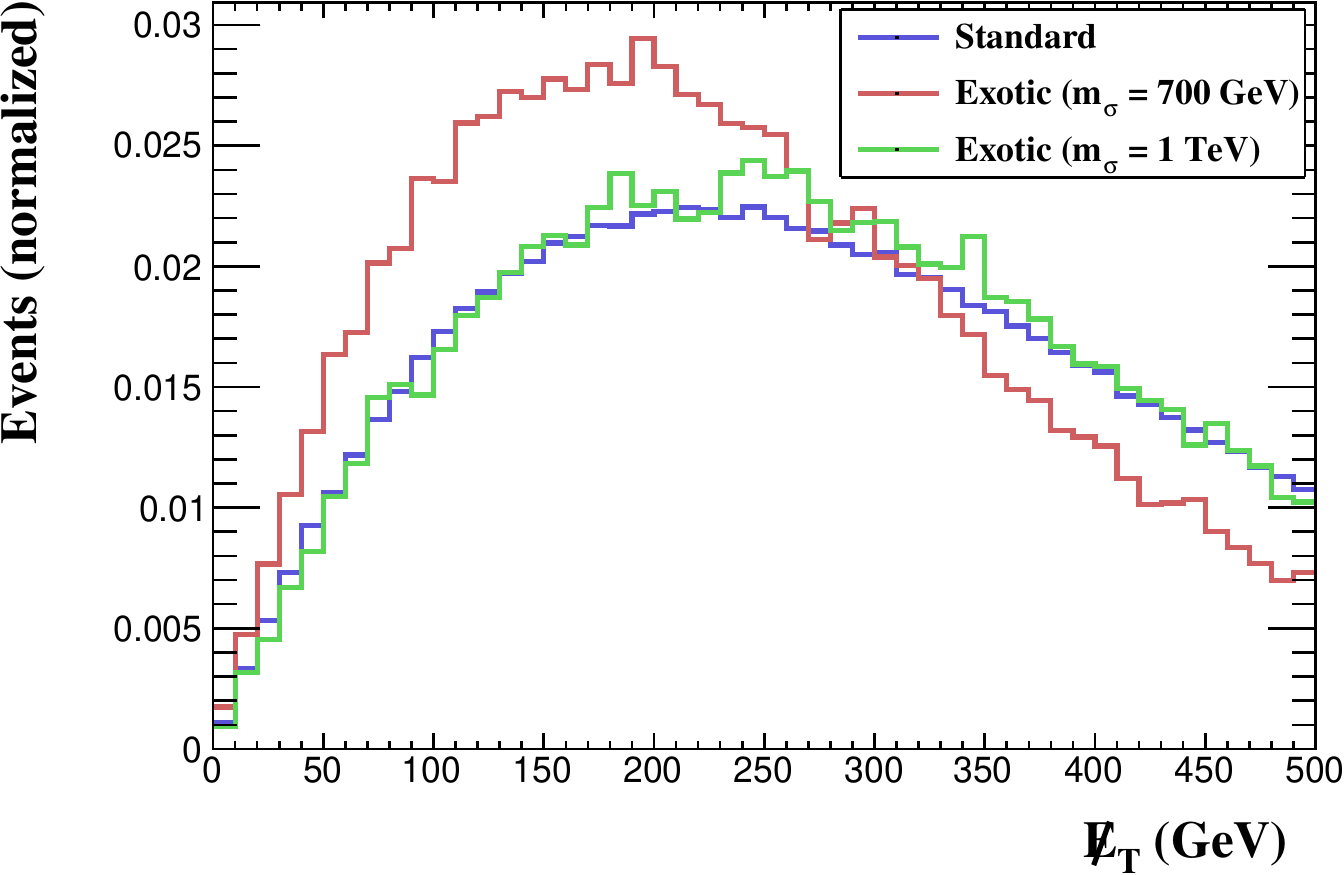}\\
\includegraphics[width=0.49\textwidth]{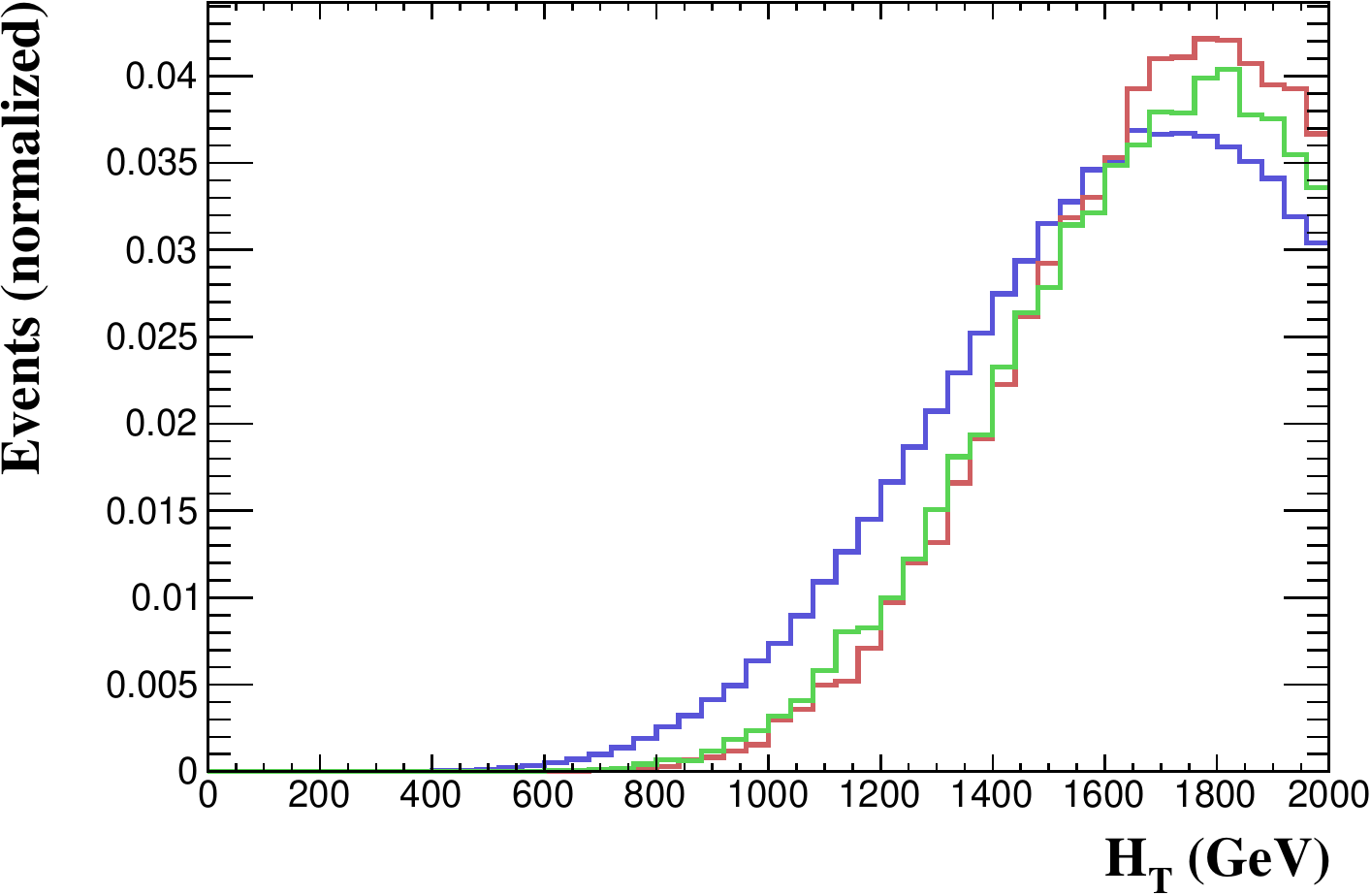}\\
\includegraphics[width=0.49\textwidth]{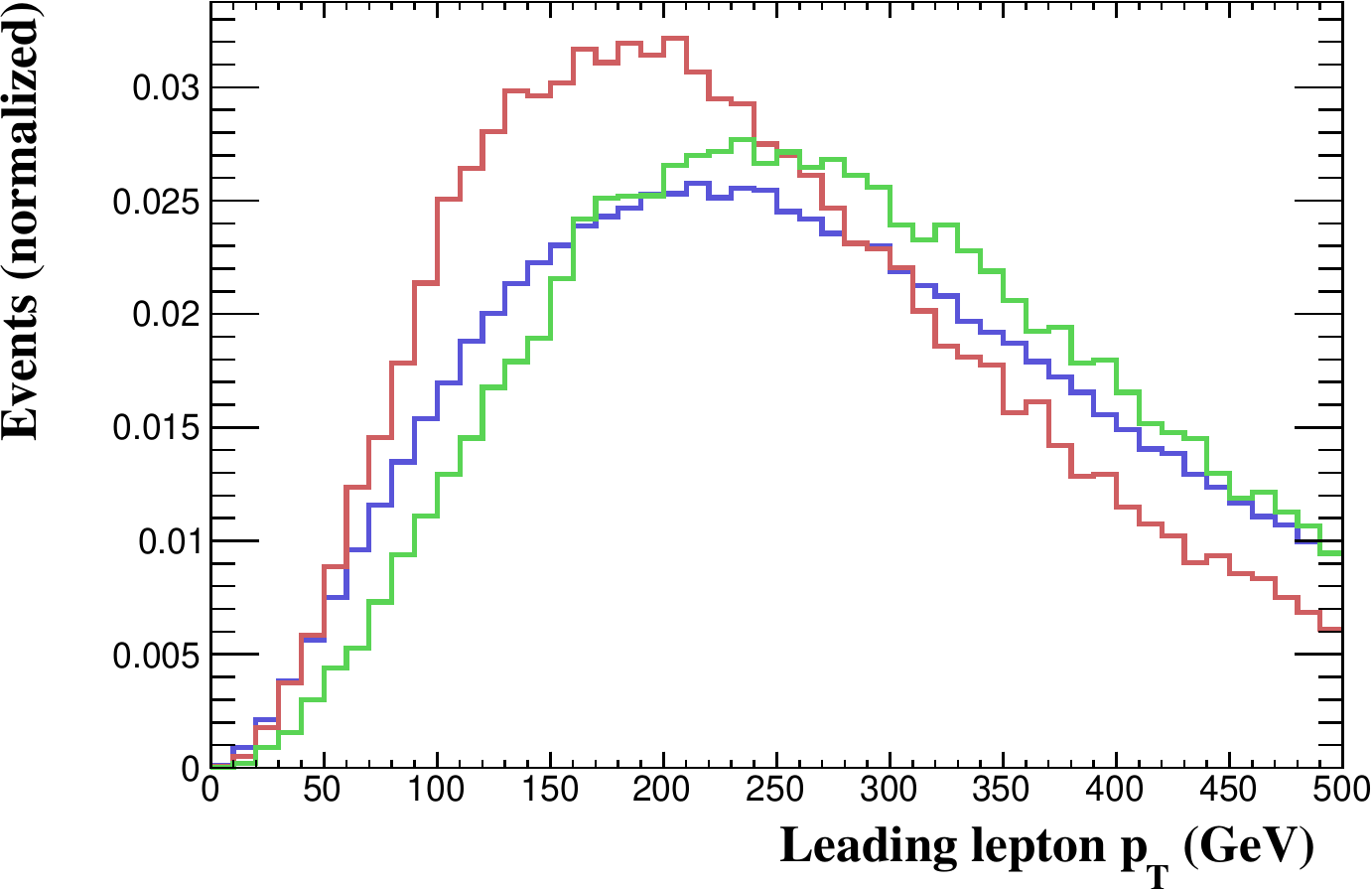}
\end{tabular}
 \caption{\em Missing transverse energy $\etmiss$ (upper panel), transverse energy $H_T$ (middle panel) and leading lepton $p_T$ (lower panel) normalised kinematic 
 distributions (for $m_B = 1.2$~TeV) for the process Eq.~\eqref{ec:BBWt} (blue) and the process $p p \to B\bar{B} \to \sigma b \sigma \bar b$ 
 with $\sigma \to W^+ W^-$, for $m_\sigma = 700$ GeV (red) and $m_\sigma = 1$ TeV (green).} 
\label{fig:MET}
\end{figure}

For illustration, Fig.~\ref{fig:MET} presents the normalised
kinematical distributions of the missing transverse energy $\etmiss$, transverse energy $H_T$ and leading lepton transverse momentum ($p_T$), for the process 
in Eq.~\eqref{ec:BBWt} with a standard VLQ decay mode ($B \to W t$) and for the process $p p \to B \bar B \to \sigma b \, \sigma \bar b$ with $\sigma \to W^+ W^-$, 
respectively for $m_\sigma = 700$ GeV and $m_\sigma = 1$ TeV, as obtained at parton level using {\tt Madgraph$\_$aMC@NLO}~\cite{Alwall:2014hca} 
and {\tt MadAnalysis~5}~\cite{Conte:2012fm}. 
The similarity of the kinematical distributions make it apparent that, 
despite being designed for a specific decay mode, the ATLAS search~\cite{Aaboud:2018xpj} is flexible enough to provide limits on 
other non-standard decay modes of VLQ.
In addition, we note that 
this search is also sensitive to single $T$ production followed by the exotic decay $T \to \sigma t$ ($\sigma \to W^+ W^-$):
\begin{equation}
pp \to T b j \,,\quad T \to \sigma t \to W^+ W^- W^+ b \,,
\end{equation}
since two same-sign or three leptons can be produced from the decay of a single $T$ state. 
Therefore, the same inclusive search can be used to probe both the mass and the coupling of new $T$ quarks with exotic decays.

\vspace{2mm}

\noindent 2. {\em ATLAS search for opposite-sign dileptons or three leptons plus $b$-tagged jets~\cite{Aaboud:2018saj}.}
This analysis focuses on the decays of vector-like quarks that involve a $Z$ boson,
\be
\begin{aligned}
pp &\to T + X \to Zt + X \,,  \\
pp &\to B + X \to Bb + X \,,
\end{aligned}
\ee
with $X$ being additional particles, that is, another heavy VLQ in pair production or light SM quarks in single production. The event selection 
requires one same-flavour opposite-sign lepton pair with an invariant mass consistent with $M_Z$, as well as $b$-tagged jets and large $H_T$ (now 
defined with jets only). The analysis defines several signal regions according to the number of large-radius jets present with large mass.  
This search is sensitive to single and pair production of $T$ or $B$ quarks followed by the exotic decays in Eq.~\eqref{ec:TBexo} and $\sigma \to ZZ$, with 
the advantage of the combinatorial factor stemming from the two $Z$ bosons that can decay leptonically. 

\vspace{2mm}

\noindent 3. {\em ATLAS search for multi-jets with several $b$-tagged jets together with one charged lepton or large $\etmiss$~\cite{Aaboud:2018xuw}.}
This analysis focuses on pair production
\begin{align}
pp \to T \bar T \to h \overset{(-)}{t} + X \,,
\end{align}
that is, when either of the heavy quarks decays into $H \overset{(-)}{t} $ and the other one decays via any of the three standard VLQ decay modes in Eq.~\eqref{ec:TBstd}. 
Besides identifying top and Higgs candidates from the reconstructed objects, a large effective mass $m_\text{eff} = H_T + \etmiss$ is required. This event 
selection is rather sensitive to pair production of $T$ states followed by the decay $T \to \sigma t$ with $\sigma \to hh$, where there are two 
Higgs bosons instead of only one as in the standard VLQ decay $T \to ht$.

With these examples one can easily understand that event selections that require high $H_T$ or $m_\text{eff}$ (characteristic of the production of 
heavy particles) and leptons or $b$ quarks are sensitive to --- but not optimised for --- the new VLQ decay modes discussed in this paper. The 
results of these searches can be reinterpreted by relaxing the conditions
\be
\begin{aligned}
& \text{Br}(T \to Wb) + \text{Br}(T \to Zt) + \text{Br}(T \to ht) = 1 \,,  \\
& \text{Br}(B \to Wt) + \text{Br}(B \to Zb) + \text{Br}(B \to hb) = 1 \,,
\end{aligned}
\ee
and introducing the additional decays in Eq.~(\ref{ec:TBexo}). On the other hand, there are searches that attempt to reconstruct the heavy quark 
masses by assuming a particular decay chain and kinematics~\cite{Aaboud:2017zfn,Aaboud:2018uek}. Their sensitivity to the new modes in which 
either there are additional particles (e.g. in $T$ decays) or the kinematical configuration is different (in $B$ decays) is degraded.

\section{Conclusions}
\label{sec:5}

Extensions of the SM that include new VLQs usually include extra new particles that may mediate new production or decay mechanisms for the VLQs. 
In this paper, the focus is the study of the phenomenology of the Minimal Linear $\sigma$ Model, that is a renormalisable extension of the SM based on the global spontaneous symmetry breaking $SO(5)\to SO(4)$ that improves on several limitations of traditional Composite Higgs models. 
The model includes several VLQs in the spectrum, as well as a new scalar $\sigma$ that can mediate novel decay channels for the VLQs.

A scan over the allowed parameter space shows that the new decay modes $T \to \sigma t$, $B \to \sigma b$ can have large 
branching ratios, and even dominate over the  `standard' decays in Eqs.~(\ref{ec:TBstd}). Moreover, the single production of $T$ quarks can be sizable and dominate over pair production, as a consequence of the interplay among the new VLQ states and the third generation SM quarks. This is a salient feature of realistic models (as opposed to simplified VLQ models), not generally possible with just one VLQ state or multiplet.

The pair and single productions of VLQs lead final states with multiple bosons, e.g.~$T \bar T \to \sigma t \sigma \bar t \to W^+ W^- W^+ b \, W^+ W^- W^- \bar b$.
These new decay modes can in principle be captured by the current searches at the LHC. In this sense, the observability of such new 
decay modes requires that VLQ searches must be rather generic, i.e. requiring the presence of multiple $b$ quarks, $W/Z/h$ bosons, high transverse energy, 
etc, and a few examples of ATLAS and CMS searches, which are sensitive to the new decay modes, have been discussed in this work, and 
can be straightforwardly interpreted to set limits on those decays.

To conclude, the possibility of new VLQ decays is open and demands more generic experimental analyses, with a wide sensitivity beyond 
the `standard' decays, and providing limits on the new modes. Dedicated searches for these new modes would benefit from the conspicuous multi-boson 
signatures produced in these decays, and would also be welcome.


\section*{Acknowledgments}

J.A.G. and L.M. thanks Stefano Rigolin for useful discussions.
The authors acknowledge partial financial support by the Spanish MINECO through the Centro de excelencia Severo Ochoa 
Program under grant SEV-2016-0597. J.A.G., L.M. and J.A.A.S. acknowledge partial financial support by the Spanish ``Agencia Estatal de Investigaci\'on'' 
(AEI) and the EU ``Fondo Europeo de Desarrollo Regional'' (FEDER) through the projects FPA2016-78645-P and FPA2016-78220-C3-1-P.
L.M. acknowledges partial financial 
support by the Spanish MINECO through the ``Ram\'on y Cajal'' programme (RYC-2015-17173). 
J.M.N. acknowledges support from the Ram\'on y Cajal Fellowship contract RYC-2017-22986 and 
from the Spanish Proyectos de I$+$D de Generaci\'on de Conocimiento via grant PGC2018-096646-A-I00.
L. M. and J.M.N. also acknowledge support from the 
European Union's Horizon 2020 research and innovation programme under the Marie Sklodowska-Curie grant agreements 690575  (RISE InvisiblesPlus) and 
674896 (ITN ELUSIVES).

\appendix
\begin{widetext}
\boldmath
\section{Expressions for the corrections to $S$ and $T$ parameters}
\unboldmath
\label{sec:a}

The explicit expression for $\Delta T^{(h \ \text{and} \ \sigma)}$ is given by\footnote{This expression differs from the one in Eq.~(51) of \cite{Feruglio:2016zvt}: an $\alpha$ is missing multiplying $\Delta T^{(h \ \text{and} \ \sigma)}$ in the LH side of that expression, while a minus sign should be present in the right-hand side.}
\be
\begin{aligned}
\Delta T^{(h \ \text{and} \ \sigma)}=&\phantom{+}\frac{3 s^2_\gamma}{16s^2_W\pi}
\bigg[
m^2_h\frac{\log\left(m^2_h/m^2_W\right)}{m^2_W-m^2_h}
-m^2_\sigma\frac{\log\left(m^2_\sigma/m^2_W\right)}{m^2_W-m^2_\sigma}+ \\
&\hspace{2cm}+\frac{m^2_Z}{m^2_W}
\left(-m^2_h\frac{\log\left(m^2_h/m^2_Z\right)}{m^2_Z-m^2_h}
+m^2_\sigma\frac{\log\left(m^2_\sigma/m^2_Z\right)}{m^2_Z-m^2_\sigma}
\right)
\bigg].
\end{aligned}
\ee
On the other side, for the computation of $\Delta S^{(h \ \text{and} \ \sigma)}$, the explicit expression of $S_\text{SM}^h$ can be found in Refs.~\cite{Novikov:1992rj,Orgogozo:2012ct}\footnote{A typo is present in Eq.~(F.7) of Ref.~\cite{Novikov:1992rj} that has been corrected in Ref.~\cite{Novikov:1995vu}.}:
\begin{equation}
S_\text{SM}^h\left(x\right)=
\frac{1}{\pi}\left[\frac{x}{12\left(x-1\right)}\log\left(x\right)
+\left(-\frac{x}{6}+\frac{x^2}{12}\right)F\left(x\right)
-\left(1-\frac{x}{3}+\frac{x^2}{12}\right)F'\left(x\right)\right]\,,
\end{equation}
where $x\equiv m^2/m^2_Z$, and with the $F(x)$ and $F'(x)$ functions given by
\be
\begin{aligned}
F\left(x\right)&=1+\left(\frac{x}{x-1}-\frac{x}{2}\right)\log\left(x\right)-x\sqrt{\frac{4}{x}-1}\arctan\left(\sqrt{\frac{4}{x}-1}\right)\,, \\
F'\left(x\right)&=-1+\frac{x-1}{2}\log\left(x\right)+(3-x)\sqrt{\frac{x}{4-x}}\arctan\left(\sqrt{\frac{4}{x}-1}\right)\,,
\end{aligned}
\ee
for $x<4$, whereas for $x>4$ \footnote{Eq.~(55) of Ref.~\cite{Feruglio:2016zvt} disagrees for  a minus sign with respect to Refs.~\cite{Novikov:1995vu,Orgogozo:2012ct}. The correct version is in Eq.~\eqref{eq:S_param_corrected} that matches Refs.~\cite{Novikov:1995vu,Orgogozo:2012ct}.}
\be
\begin{aligned}
F\left(x\right)&=1+\left(\frac{x}{x-1}-\frac{x}{2}\right)\log\left(x\right)+x\sqrt{1-\frac{4}{x}}\log\left(\sqrt{\frac{x}{4}-1}+\sqrt{\frac{x}{4}}\right)\,,\\
F'\left(x\right)&=-1+\frac{x-1}{2}\log\left(x\right)+(3-x)\sqrt{\frac{x}{x-4}}\log\left(\sqrt{\frac{x}{4}-1}+\sqrt{\frac{x}{4}}\right)\,.
\end{aligned}
\label{eq:S_param_corrected}
\ee

The generalized contribution to the $T$ parameter from VLQs with arbitrary couplings was presented at Ref.~\cite{Anastasiou:2009rv}:

\be
\begin{aligned}
T^f=\frac{N_c}{16\pi s^2_W c^2_W}
\bigg(\sum_{i,j}
\left[
\left(\left\vert V_L^{ij}\right\vert^2+\left\vert V_R^{ij}\right\vert^2\right)\theta_+\left(x_i,x_j\right)+2\text{Re}\left(V_L^{ij}V_R^{ij\ast}\right)\theta_-\left(x_i,x_j\right)\right]+ \\
-\frac{1}{2}\sum_{i,j}
\left[
\left(\left\vert C_L^{ij}\right\vert^2+\left\vert C_R^{ij}\right\vert^2\right)\theta_+\left(x_i,x_j\right)+2\text{Re}\left(C_L^{ij}C_R^{ij\ast}\right)\theta_-\left(x_i,x_j\right)
\right]
\bigg),\label{eq:T_ferm_generic}
\end{aligned}
\ee
where $N_c$ is the number of colours, $x_i\equiv m^2_i/m^2_Z$ with $m_i$ the VLQ masses, and $V_{L,R}$ and $C_{L,R}$ are the matrices given in Eqs.~\eqref{VLRdefinition} and \eqref{CLRdefinition}. The $\theta_\pm$ functions are defined as
\begin{align}
\theta_+\left(x_1,x_2\right)&\equiv x_1+x_2-\frac{2x_1x_2}{x_1-x_2}\log\left(\frac{x_1}{x_2}\right)
-2\left(x_1\log\left(x_1\right)+x_2\log\left(x_2\right)\right)+\frac{x_1+x_2}{2}\Delta\nn \\
\theta_-\left(x_1,x_2\right)&\equiv2\sqrt{x_1x_2}\left(\frac{x_1+x_2}{x_1-x_2}\log\left(\frac{x_1}{x_2}\right)-2+
\log\left(x_1x_2\right)-\frac{\Delta}{2}\right)\,,
\end{align}
where $\Delta$ is a divergent quantity arising in the dimensional regularisation. This contribution disappears in the sum of all the contributions.

Similarly for the $S$ parameter, the generic expression can be found in Ref.~\cite{Carena:2006bn} and reads
\be
S^{f}=\frac{N_c}{2\pi}
\bigg(\sum_{i,j}
\bigg[
\left(
X_L^{ij\ast} Y^{ij}_L+X_R^{ij\ast} Y^{ij}_R\right)\chi_{+}\left(\widehat\cM_{ii},\widehat\cM_{jj}\right)+\left(
X_L^{ij\ast} Y^{ij}_R+X_R^{ij\ast} Y^{ij}_L\right)\chi_{-}\left(\widehat\cM_{ii},\widehat\cM_{jj}
\right)
\bigg]\bigg)\,,
\label{eq:S_ferm_generic}
\ee
where here $X_{L,R}$ are the isospin-dependent coupling matrices in the mass basis (that is, the $U_{L,R} \ \mathcal{C}_{L,R} \ U^\dagger_{L,R}$ matrices present at Eq.~\eqref{CLRdefinition}), the $Y_{L,R}$ are the hypercharge matrices of the fermions and the $\chi_\pm$ functions are defined as
\be
\begin{aligned}
\chi_+\left(m_1,m_2\right)=&
\frac{5\left(m_1^4+m_2^4\right)-22m_1^2m_2^2}{9\left(m_1^2-m_2^2\right)^2}+\frac{3m_1^2m_2^2\left(m_1^2+m_2^2\right)-\left(m_1^6+m_2^6\right)}{3\left(m_1^2-m_2^2\right)^3}\log\left(\frac{m_1^2}{m_2^2}\right)-\frac{2}{3}\log\left(\frac{m_1m_2}{\mu^2}\right)\,,\\
\chi_-\left(m_1,m_2\right)=&
\frac{m_1 m_2}{\left(m_1^2-m_2^2\right)^3}\left[m_1^4-m_2^4-2m_1^2m_2^2\log\left(\frac{m_1^2}{m_2^2}\right)\right]\,,
\end{aligned}
\ee
being $\mu$ an arbitrary scale that disappears as long as $\text{Tr}\left[X^\dagger_LY_L+X^\dagger_RY_R\right]=0$.

In the \MLsM, the hypercharge matrices are defined by
\be
Y_{L,R}=U_{L,R} \, \cY_{L,R} \, U^\dagger_{L,R}\,,
\ee
where the $\cY_{L,R}$ are given by
\be
\cY_{L,R}=\diag\left(\frac{7}{6},\cY^{\mathcal{T}}_{L,R},\cY^{\mathcal{B}}_{L,R},-\frac{5}{6}\right)\,,
\ee
with
\be
\begin{aligned}
\cY^{\mathcal{T}}_{L}&=\diag\left(\frac{1}{6},\frac{1}{6},\frac{7}{6},\frac{2}{3},\frac{2}{3},\frac{1}{6}\right)\,,\qquad
&\cY^{\mathcal{T}}_{R}&=\diag\left(\frac{2}{3},\frac{1}{6},\frac{7}{6},\frac{2}{3},\frac{2}{3},\frac{1}{6}\right)\,, \\
\cY^{\mathcal{B}}_{L}&=\diag\left(\frac{1}{6},\frac{1}{6},-\frac{5}{6},-\frac{1}{3},-\frac{1}{3},\frac{1}{6}\right)\,, \qquad
&\cY^{\mathcal{B}}_{R}&=\diag\left(-\frac{1}{3},\frac{1}{6},-\frac{5}{6},-\frac{1}{3},-\frac{1}{3},\frac{1}{6}\right)\,.
\end{aligned}
\ee

\section{Expressions for the partial widths}
\label{sec:b}

The explicit computation of the decay rates leads to the following results: for the process involving a scalar,
\begin{align}
\Gamma_{\psi_j\rightarrow \psi_i s }=&\phantom{+}
\frac{1}{32\pi}
\frac{\lambda^{1/2}\left(m^2_j,m_i^2,m_s^2\right)}{m_j}
\bigg[\left(\left(g^{\psi s}_L\right)_{ij}^2+\left(g^{\psi s}_R\right)_{ij}^2\right)\left(1+\frac{m_i^2-m^2_s}{m_j^2}\right)+4\left(g^{\psi s}_L\right)_{ij}\left(g^{\psi s}_R\right)_{ij}\frac{m_i}{m_j}\bigg]\,, 
\label{eq:Decay_Qj_into_Qi_and_S}
\end{align}
with $s$ being either $h$ or $\sigma$, $\psi_{i,j}$ being components of $\widehat\cT$ or $\widehat\cB$ of Eqs.~\eqref{eq:phys_states}, the couplings $g^{\psi s}_{L,R}$ defined in Tab.~\ref{tab:Decay_couplings} and $\lambda$ given by:
\begin{align}
\lambda=\Big[m_j^2-\left(m_i^2+m_s^2\right)^2\Big]\Big[m_j^2-\left(m_i^2-m_s^2\right)^2\Big];
\label{eq:Decay_Qj_into_Qi_and_S}
\end{align}
instead, for a process with a gauge boson,
\begin{multline}
\Gamma_{\psi_j\rightarrow \psi'_i V}=\phantom{+}
\frac{g^2}{64\pi}k_V
\frac{m_j}{m_V^2}\lambda^{1/2}\left(m^2_j,m_i^2,m_V^2\right)\times \\
\times\bigg[
\left(\left(g_L\right)_{ij}^2+\left(g_R\right)_{ij}^2\right)
\left(1+\frac{m_V^2-2m_i^2}{m_j^2}+\frac{m_i^4+m_i^2m_V^2-2m_V^4}{m_j^4}\right)
-12\left(g_L\right)_{ij}\left(g_R\right)_{ij}\frac{m_{i}m_V^2}{m^3_{j}}
\bigg],
\label{eq:Decay_Qj_into_Qi_and_V}
\end{multline}
with $k_V=1$ for $V=W^\pm$ and $k_V=1/\left(2c_W^2\right)$ for $V=Z$ and the couplings given again at Tab.~\ref{tab:Decay_couplings}. 

\end{widetext}

\begin{center}
\begin{table}[h!]
\centering
 \begin{tabular}{|c|c|}
\hline
& \\[-3mm]
$\bm{\psi_j\rightarrow \psi'_i X}$ & $\bm{\left(g_{L,R}\right)_{ij}}$  \\[1mm]
\hline
& \\[-3mm]
$\psi_j\rightarrow \psi_i\, s$ & $\left(g^{\psi s}_L\right)_{ij}=\left(U^\psi_R\, \mathpzc{Y}_s^{\psi \dagger}\, U^{\psi\dagger}_L \right)_{ij}$ \\
& $\left(g^{\psi s}_R\right)_{ij}=\left(U^\psi_L\, \mathpzc{Y}_s^\psi\, U^{\psi\dagger }_R\right)_{ij}$ \\[3mm]
\hline
& \\[-3mm]
$\widehat\cT_j\rightarrow K^u\, W^- $ & $\left(g_{L,R}\right)_{ij}=\left( U^{\cT\dagger}_{L,R}\right)_{3\,j} $ \\[3mm]
$\widehat\cB_j\rightarrow K^{\prime d}\,W^+ $ & $\left(g_{L,R}\right)_{ij}=\left( U^{\cB\dagger}_{L,R}\right)_{3\,j} $ \\[3mm]
\hline
& \\[-3mm]
$\widehat\cT_j\rightarrow \widehat\cB_i W^+$ & $\left(g_{L,R}\right)_{ij}=\left(U^{\mathcal{B}}_{L,R}\,\cV^{\cT\cB}_{L,R}\,U^{\cT\dag}_{L,R}\right)_{ij}$ \\[3mm]
\hline
& \\[-3mm]
$\psi_j \rightarrow \psi_i\, Z $ & $\left(g_{L,R}\right)_{ij}=\left(C_{L,R}\right)_{ij}$ \\[3mm]
\hline
\end{tabular}
\caption{\em Couplings entering the decays rates in Eqs.~\eqref{eq:Decay_Qj_into_Qi_and_S} and \eqref{eq:Decay_Qj_into_Qi_and_V}.}
\label{tab:Decay_couplings}
\end{table}
\end{center}


\providecommand{\href}[2]{#2}\begingroup\raggedright\endgroup

\end{document}